\documentclass[a4paper,superscriptaddress,twocolumn,aps,prb,floatfix,citeautoscript]{revtex4-1}
\usepackage{times} 
\usepackage{graphicx}
\usepackage{color}
\usepackage{amsmath}
\usepackage{amssymb}
\usepackage[english]{babel}
\usepackage[utf8]{inputenc}
\usepackage{bm}
\usepackage{url}
\usepackage[colorlinks=true,linkcolor=blue,citecolor=blue,urlcolor=blue]{hyperref}

\newcommand{\I}{\text{i}}
\newcommand{\ket}[1]{\left| #1 \right>} 
\newcommand{\vect}[1]{\mathbf{#1}}
\newcommand{\be}{\begin{equation}}
\newcommand{\ee}{\end{equation}}
\newcommand{\bea}{\begin{eqnarray}}
\newcommand{\eea}{\end{eqnarray}}

\newcommand{\ba}{\begin{array}}
\newcommand{\ea}{\end{array}}

\newcommand{\xc}{\text{xc}}

\newcommand{\wpf}{$W$-PF }

\newcommand{\cre}[1]{\hat{c}^\dagger_{#1}}
\newcommand{\an}[1]{\hat{c}_{#1}}

\begin{document}
\title{Introducing screening in one-body density matrix functionals: impact on the Extended Koopmans' Theorem's charged excitations of model systems}

\newcommand{\lcpq}{Laboratoire de Chimie et Physique Quantiques, Universit\'e de Toulouse, CNRS, UPS, France and European Theoretical Spectroscopy Facility (ETSF)}
\newcommand{\lpt}{Laboratoire de Physique Th\'eorique, Universit\'e de Toulouse, CNRS, UPS, France and European Theoretical Spectroscopy Facility (ETSF)}
\author{S. Di Sabatino}
\affiliation{\lcpq}
\affiliation{\lpt}
%
\author{J. Koskelo}
\affiliation{\lpt}
%
\author{J.~A. Berger}
\affiliation{\lcpq}
%
%
\author{P. Romaniello}
\affiliation{\lpt}
%


\begin{abstract}
In this work we get insight into the impact of reduced density matrix functionals on the quality of removal/addition energies obtained using the Extended Koopmans' Theorem (EKT).  
Within  reduced density matrix functional theory (RDMFT) the EKT approach reduces to a matrix diagonalization, whose ingredients are the one- and two-body reduced density matrices. A striking feature of the EKT within RDMFT is that it opens a band gap, although too large, in strongly correlated materials, which are a challenge for state-of-the-art methods such as $GW$. 
Using the one-dimensional Hubbard model and the homogeneous electron gas as test cases, we find that: i) with exact or very accurate density matrices the EKT systematically overestimates the band gap in the Hubbard model and the bandwidth in the homogeneous electron gas; ii) with approximate density matrices, instead, the EKT can benefit from error cancellation. In particular we test a new approximation which combines RPA screening with the Power functional (PF) approximation to the two-body reduced density matrix introduced by Sharma et al. [Phys. Rev. B 78, 201103(R) (2008)]. An important feature of this approximation is that it reduces the EKT band gap in the studied models; it can hence be a promising approximation for correcting the EKT band-gap overestimation in strongly correlated materials.
\end{abstract}
\date{\today}
\maketitle

\section{Introduction}
The Extended Koopmans' Theorem (EKT) \cite{morrell_JCP1975,smith_JCP1975} offers an interesting tool for the calculation of removal/addition energies from any level of theory.\cite{kent_PRB1998,
doi:10.1021/acs.jctc.1c00100,
pernal_CPL2005,
Leiva200645} In particular within reduced density matrix functional theory (RDMFT) \cite{PhysRev.97.1474,PhysRevB.12.2111,Pernal_TOPCURRCHEM2015}, the EKT approach is based on a matrix diagonalization, whose ingredients are the one- and two-body reduced density matrices (1-RDM and 2-RDM, respectively).
This formulation is particularly appealing, because it does not rely on the knowledge of the ground-state many-body wavefunction of the $N$-electron system, but on simpler quantities, namely the natural orbitals and occupation numbers, i.e., the eigenvectors and eigenvalues of the 1-RDM. Within RDMFT, indeed, the one-body reduced density matrix, thanks to a one-to-one map with the ground-state many-body wavefunction, can give access to all ground-state observables of the system, provided that their functional expression in terms of the 1-RDM is known. In particular the total energy is a functional of the 1-RDM and its minimization under a set of physical constraints (ensemble $N$-representable constraints) gives the exact 1-RDM. In practice the electron-electron interaction energy, which can be  expressed in terms of the 2-RDM, is an unknown functional (more precisely its correlation part) of the 1-RDM, and approximations are needed. 
The EKT offers a path towards the description of photoemission in strongly correlated materials, which is a challenge for \textit{ab initio} theories. We have indeed shown that EKT energies within the so-called diagonal approximation (DEKT) \cite{frontiers_2021} are equivalent to the energies obtained within the many-body effective energy theory (MEET) \cite{stefano} at its lowest order approximation in terms of the 1-RDM and 2-RDM. At this level of approximation and within RDMFT the MEET gives a qualitatively good description of the photoemission spectra of several paramagnetic transition-metal oxides, which are insulators, unlike mean-field theories and the more advanced $GW$ method, which describe them as metals \cite{stefano,stefano_JCTC,stefano_PRR2021}. The band gap, however, is largely overestimated. Indeed, the EKT/MEET (at its lowest order approximation) tends to overestimate 
the fundamental band gap with a magnitude which depends on the degree of correlation of the system under study.  However this overestimation can be amplified by commonly used approximations to the two-body density matrix employed in RDMFT. In particular we used the Power functional (PF) proposed by Sharma \textit{et al.}\cite{sharma_PRB08}, which is the only one that, to the best of our knowledge, has been used in solids, but similar trends are expected using approximations of the same type, i.e., the so-called $\mathcal{JK}$ functionals, which involve only Coulomb- ($\mathcal{J}$) and exchange- ($\mathcal{K}$) like integrals involving the natural orbitals \cite{Cioslowski_JCP2003}.Therefore in this work we propose a variation of the PF functional in which RPA screening is taken into account, referred to as screened-PF (\wpf) throughout this article. This is motivated by the fact that in many-electron systems screening becomes important and, for example, in the context of many-body perturbation theory (MBPT) based on Green's functions the improvement of the $GW$ approximation over Hartree-Fock is precisely due to the screening of the Coulomb interaction. We also consider the BBC1 functional\cite{bbc} which, as we shall see, shows important physical features in the correlation energy and the natural orbital occupation numbers, which can have an impact on the quality of the EKT removal/addition energies.  We test the quality of these approximations using the one-dimensional Hubbard model and the homogeneous electron gas (HEG) as benchmark systems. 

The paper is organized as follows. In Sec.~\ref{Sec:Theory} we give the basic equations of the EKT as well as RDMFT and we derive the \wpf. In Sec.~\ref{Sec:Models} we describe the two models used. Computational details are discussed in Sec.~\ref{Sec:Computation}. In Sec.~\ref{Sec:Results} we report and discuss our results. In  Sec.~\ref{Sec:Conclusions} we draw our conclusions and perspectives.

\section{Theoretical framework\label{Sec:Theory}}
\subsection{The Extended Koopman’s Theorem}
Within the EKT one considers 
the following 
wave functions 
for the $\nu$-th one-particle removal excitations \cite{kent_PRB1998}
\be
\ket{\Psi^{N-1}_\nu}=\hat{O}_\nu\ket{\Psi_0^N},
\ee
where $\ket{\Psi_0^N}$ is the $N$-particle ground-state wavefunction, $\hat{O}_\nu$
the electron annihilation operator
$
\hat{O}_\nu=\sum_i C^R_{\nu i}\hat{c}_i,
$
and $\{C^R_{\nu i}\}$ are a set of coefficients to be determined. Here the indices $i,j,..$ refer to a general basis of spinorbitals, i.e., $i=I\sigma$ comprises the orbital index $I$ and the spin $\sigma$. 
The corresponding removal energy
is given by
\begin{equation}
\epsilon^R_\nu=-\frac{\langle\Psi^N_0|\hat{O}_\nu^\dagger[\hat{H},\hat{O}_\nu]| \Psi^N_0\rangle}{\langle\Psi^N_0|\hat{O}_\nu^\dagger\hat{O}_\nu |\Psi^N_0\rangle}.
\label{Eqn:valence-energy}
    \end{equation}

The stationary condition (with respect to the coefficients $C^R_{\nu i}$ ) for $\epsilon^R_i$ leads to the following generalized eigenvalue equation
    \begin{equation}
    (\mathbf{F}^R-\epsilon^R_\nu\mathbf{S}^R)\mathbf{C}^R_\nu=0,
    \label{Eqn:removal-energy}
   \end{equation}
with $F^R_{ij}=-\langle\Psi^N_0|\hat{c}_j^\dagger[\hat{H},\hat{c}_i]| \Psi^N_0\rangle$ and $\mathbf{S}^R$ the one-body reduced density matrix $S^R_{ij}=\gamma_{ij}=\langle\Psi^N_0|\hat{c}_j^\dagger\hat{c}_i| \Psi^N_0\rangle$.
If one defines the matrix  $\mathbf{\Lambda}^R=[\mathbf{S}^{R}]^{-1}\mathbf{F}^R$ in the basis of natural orbitals, with ${S}^R_{ij}=n_i\delta_{ij}$ and works out the commutator in $F^R_{ij}$ using the  many-body Hamiltonian
$\hat{H}=\sum_{ij}h_{ij}\hat{c}_i^\dagger \hat{c}_j+\frac{1}{2}\sum_{ijkl}V_{ijkl}\hat{c}_i^\dagger\hat{c}_j^\dagger\hat{c}_l\hat{c}_k$,
one arrives at 
\begin{equation}
\Lambda^R_{ij}=\frac{1}{n_i}\left[n_ih_{ji}+\sum_{klm}V_{jmkl}
  \Gamma^{(2)}_{klmi}\right],
  \label{eqn:lambdaR}
\end{equation}
where $\Gamma^{(2)}_{klji}=\langle\Psi^N_0|\hat{c}^\dagger_i\hat{c}^\dagger_j\hat{c}_l\hat{c}_k|\Psi^N_0\rangle$ are the 2-RDM matrix elements,
$h_{ij}=\int d\mathbf{x}\phi_i^*(\mathbf{x})h(\mathbf{x})
\phi_j(\mathbf{x})$ are the matrix elements of the one-particle noninteracting Hamiltonian $h(\vect{x})=-\nabla^2/2+v_{\text{ext}}(\vect{x})$, with $v_{\text{ext}}(\vect{x})$ a static external potential, and
$V_{ijkl}=\int d\mathbf{x}d\mathbf{x}'\phi_i^*(\mathbf{x})\phi_j^*(\mathbf{x}')v(\mathbf{x},\mathbf{x}')\phi_k(\mathbf{x})\phi_l(\mathbf{x})$ are the matrix elements of the Coulomb interaction $v(\vect{x},\vect{x}')$.
Diagonalization of $\mathbf{\Lambda}^R$ yields the removal energies $\epsilon_\nu^R$ as eigenvalues.\citep{morrell_JCP1975,pernal_CPL2005} The diagonal elements of $\mathbf{\Lambda}^R$ are referred in literature as the energies of the EKT within the diagonal approximation (DEKT).\cite{kent_PRB1998}

Similar equations hold for the addition energies. One can indeed start from the wave function 
$\ket{\Psi^{N+1}_\nu}=\hat{O}^\dagger_\nu\ket{\Psi_0^N}$, with $
\hat{O}^\dagger_\nu=\sum_i C^A_{\nu i}\hat{c}^\dagger_i$,
write the addition energy $\epsilon^A_\nu$ as
\begin{equation}
\epsilon^A_\nu=\frac{\langle\Psi^N_0|[\hat{H},\hat{O}_\nu]\hat{O}_\nu^\dagger| \Psi^N_0\rangle}{\langle\Psi^N_0|\hat{O}_\nu\hat{O}_\nu^\dagger|\Psi^N_0\rangle }
\label{Eqn:conduction-energy}
    \end{equation}
    and in a similar way as for $\epsilon^R_\nu$ we arrive at the generalized eigenvalue equation
 \begin{equation}
    (\mathbf{F}^A-\epsilon^A_\nu\mathbf{S}^A)\mathbf{C}^A_\nu=0,
    \label{Eqn:addition-energy}
   \end{equation}   
with  $F^A_{ij}=\langle\Psi^N_0|\hat{c}_i[\hat{H},\hat{c}_j^\dagger]| \Psi^N_0\rangle$ and $\mathbf{S}^A$ related to the one-body density matrix as $S^A_{ij}=1-\gamma_{ij}$.  
Similarly to the removal energy problem, using the basis of natural orbitals, one can work out the commutator in $F^A_{ij}$ and reformulate the problem in terms of the matrix  $\mathbf{\Lambda}^A=[\mathbf{S}^{A}]^{-1}\mathbf{F}^A$ \footnote{Since in the basis of natural orbitals the $S^{R}$ ($S^{A}$) matrix is a diagonal matrix with the natural occupation numbers $n_i$ (1-$n_i$) as elements, the invertibility of this matrix is strictly related to the non-existence of so-called pinned states, i.e. states with occupation numbers equal to 1 or 0.  This is an important question that has several consequences.\cite{Giesbertz_JCP2013, Baldsiefen_PRA2015} Here we assume that $S^{R}$ ($S^{A}$) is invertible in a restricted space (of natural orbitals) in which the corresponding KS orbitals are occupied (unoccupied). This is a reasonable assumption.}, which reads
\begin{eqnarray}
\Lambda^A_{ij}&=&\frac{1}{(1-n_i)}\times\nonumber\\
&&\left[(1-n_i)h_{ji}+\sum_k \left(V_{jkik} -V_{jkki}\right)n_k -\sum_{klm}V_{jmkl}\Gamma^{(2)}_{klmi}\right].\nonumber\\
\label{Eqn:Lagrangian_A}
\end{eqnarray}
Diagonalization of $\mathbf{\Lambda}^A$ yields the addition energies $\epsilon_\nu^A$ as eigenvalues.
The EKT approach offers a way to build approximations for the spectral function\cite{frontiers_2021,doi:10.1021/acs.jctc.1c00100,Sharma13}.
In the basis of natural orbitals and within the DEKT the approximate spectral function assumes a particular simple form given by
\be
A(\omega)=\sum_i \left[n_i\delta(\omega-\epsilon_i^R)+(1-n_i)\delta(\omega-\epsilon_i^A)\right].
\label{eqn:spectral}
\ee

\subsection{RDMFT}
In RDMFT the ground-state total energy is a unique functional of the 1-RDM
\begin{eqnarray}
E[\gamma]&=&\int d\mathbf{x}d\mathbf{x}'\delta(\mathbf{x}-\mathbf{x}')h(\mathbf{x})\gamma(\mathbf{x},\mathbf{x}')\nonumber\\
&&+\frac{1}{2}\int d\mathbf{x}d\mathbf{x}'v(\mathbf{x},\mathbf{x}')\Gamma^{(2)}[\gamma](\mathbf{x},\mathbf{x}';\mathbf{x},\mathbf{x}'),
\end{eqnarray} 
where the 2-RDM can be factorized as
\begin{eqnarray}
 \Gamma^{(2)}[\gamma](\mathbf{x},\mathbf{x}^\prime; \mathbf{x},\mathbf{x}^\prime)&=&
\gamma(\mathbf{x},\mathbf{x})
 \gamma(\mathbf{x}^\prime, \mathbf{x}^\prime)- \gamma(\mathbf{x},\mathbf{x}^\prime)
 \gamma(\mathbf{x}^\prime, \mathbf{x})\nonumber\\
  &&+\Gamma^{(2)}_{\mathrm{c}}[\gamma](\mathbf{x},\mathbf{x}^\prime; \mathbf{x}, \mathbf{x}^\prime).
 \label{Eqn:Gamma_2}
\end{eqnarray}
The first and second terms on the right-hand side of Eq.~\eqref{Eqn:Gamma_2} give rise to the Hartree and exchange contributions to the total energy, whereas the last term yields the correlation energy, which is the only part unknown and which needs to be approximated. Most of the commonly used approximations are implicit functionals of the 1-RDM and explicit functionals of the natural orbitals ($\phi_i$) and occupation numbers ($n_i$), which are the eigenvectors and eigenvalues, respectively, of the 1-RDM (i.e., $\gamma(\mathbf{x},\mathbf{x}^\prime)=\sum_{i}^{}n_{i}\phi_{i}(\mathbf{x})\phi^*_i(\mathbf{x}^\prime))$. In particular here we focus on the $\mathcal{JK}$-only functionals, which, in their simplest form, read
\begin{eqnarray}
\Gamma^{(2)}[\gamma](\mathbf{x},\mathbf{x}^\prime;\mathbf{x},\mathbf{x}^\prime)&\approx& \sum_{ij}n_in_j\phi^*_i(\mathbf{x})\phi^*_j(\mathbf{x}^\prime)\phi_i(\mathbf{x})\phi_j(\mathbf{x}^\prime)\nonumber\\
&-&\sum_{ij}f(n_i,n_j)\phi^*_i(\mathbf{x})\phi^*_j(\mathbf{x}^\prime)\phi_j(\mathbf{x})\phi_i(\mathbf{x}^\prime),\nonumber\\
\label{eqn:2rdmJK}
\end{eqnarray}
i.e., they have the form of the Hartree-Fock exchange
modified by the function $f(n_i,n_j)$ of the occupation numbers.

In this work we will focus on the Power functional (PF) proposed by Sharma \textit{et al.},\cite{sharma_PRB08, Sharma13} which is the only one that, to the best of our knowledge, has been used in solids, for which $f^\text{PF}(n_i,n_j)=n_i^\alpha n_j^\alpha$, with $0.5\le\alpha\le 1$. Note that with $\alpha=1$ one gets the Hartree-Fock approximation to $\Gamma^{(2)}$, whereas with $\alpha=0.5$ one gets the M\"uller functional \cite{muller}. We will also employ the functional proposed by Buijse and Baerends as corrections to the M\"uller functional (BBC)\cite{bbc} for which one has to distinguish between strongly and weakly occupied orbitals. 
This distinction appears naturally when a subset of the orbitals corresponds to
occupation numbers close to 1, and the rest to occupation
numbers close to 0 (weakly correlated systems).
However, in more general situations this distinction might be an issue.
Here we will use the simplest version of the BBC functional, the BBC1,  for which
\begin{equation}
  f^{\text{BBC1}}(n_i,n_j) =
    \begin{cases}
      -\sqrt{n_in_j} & \text{if $I\neq J$ and $i$, $j$ weakly occupied} \\
      \sqrt{n_in_j} & \text{otherwise}
    \end{cases}       
\end{equation}
Extension of more advanced functionals used for finite systems to solids, such as some of the PNOF series \cite{piris1, PhysRevLett.119.063002, PhysRevLett.127.233001}, is not straightforward.

The total energy can then be expressed as a functional of $\phi_i$ and $n_i$, $E[\{n_i\},\{\phi_i\}]$;  functional minimization with respect to the natural orbitals, under orthonormality constraints, and occupation numbers, under the ensemble $N$-representability constraints ($\sum_in_i=N$, with $N$ the total number of electrons, and $0\leq n_i\leq 1$), leads to the ground-state total energy. 

\subsubsection{Screened power functional}

In the expression of the RDMFT total energy and of the EKT energies we have terms like 
\begin{equation}
\sum_{klm}V_{jmkl}\Gamma^{(2)}_{klmi}.
\end{equation}
 In the following we exploit the link between $\Gamma^{(2)}$ and the two-body Green's function $G^{(2)}$ \cite{Strinati},
$$\Gamma^{(2)}(\vect{x}_1,\vect{x}_2;\vect{x}_{1}',\vect{x}_2)=-G^{(2)}(\vect{x}_1t_1,\vect{x}_2t^+_{1};\vect{x}_{1}'t^{+++}_{1},\vect{x}_2t^{++}_{1}),$$
with $t^+_1=t_1+\delta$ ($\delta=0^+$), to get approximation to $\Gamma^{(2)}$ from approximations to the self-energy $\Sigma$ of MBPT. 
We first start from the definition of the self-energy in terms of $G^{(2)}$: 
\begin{equation}
\int d2\, \Sigma(12)G(24)=-\I\int d3 v(13)G^{(2)}(13^+;43^{++}),
\end{equation}
where $1 \equiv (\vect{x}_1, t_1)$ is a space-spin plus time composite variable.
By expressing $G$ and $G^{(2)}$ in a basis set $\{\phi_i(\mathbf{x})\}$, and by multiplying and integrating 
both sides of the equation
 with $\int d\mathbf{x}_1d\mathbf{x}_4\phi^*_m(\mathbf{x}_1)\phi_l(\mathbf{x}_4)$ we arrive at 
\begin{equation}
\sum_{i}\int dt_2 \Sigma_{mi}(t_1t_2)G_{il}(t_2t_4)=-\I \sum_{ijk}V_{mkij}G^{(2)}_{ijkl}(t_1t^+_1;t_4t^{++}_1),
\label{Eqn:general_G2}
\end{equation}
with $G_{ij}(t_1t_2)=-i\langle\Psi_0|\mathcal{T}[\an{i}(t_1)\cre{j}(t_2)] |\Psi_0\rangle$ and $G^{(2)}_{ijkl}(t_1t_2;t_3t_4)=-\langle\Psi_0|\mathcal{T}[\hat{c}_{i}(t_1)\hat{c}_{j}(t_2)\hat{c}^\dagger_{k}(t_4)\hat{c}^\dagger_{l}(t_3)]|\Psi_0\rangle$.
We now consider $t_4=t_1^{+++}$ to get $\Gamma^{(2)}$ on the right-hand side. 
Expressing the left-hand side in frequency space we arrive at
\begin{equation}
\sum_{i}\int \frac{d\omega}{2\pi}\Sigma_{mi}(\omega)G_{il}(\omega)e^{\I\omega\eta}=\I\sum_{ijk}V_{mkij}\Gamma^{(2)}_{ijkl}.
\label{Eqn:geneal_D}
\end{equation}
Approximations to the self-energy will give approximations to the term $V\Gamma^{(2)}$. In particular the frequency dependence of the self-energy is essential to have fractional occupation numbers \cite{stefano_JCP2015}, which in turn are related to the band-gap opening in strongly correlated systems \cite{stefano}. However modeling the correct frequency dependence is not easy. As a paradigmatic example we can consider the Hubbard dimer, in which the well-known $GW$ approximation to the self-energy fails to open a gap in the strongly correlated limit. \cite{stefano} We therefore consider a static self-energy, such as $GW$ with a statically screened $W$, which leads to 
\begin{equation}
\sum_{ijk}\gamma_{kj}W_{mjki}\gamma_{il}=-\sum_{ijk}V_{mkij}\Gamma^{(2)}_{\xc,ijkl},
\end{equation}
where we used the fact that $-\I \int d\omega/(2\pi) G_{il}(\omega)e^{\I\omega\eta}=\gamma_{il}$ and where we considered only the exchange-correlation contributions to $\Sigma$ and $\Gamma^{(2)}$, since the Hartree contribution to $\Gamma^{(2)}$ as functional of the 1-RDM is known. 
If we work in the basis of natural orbitals we get
\begin{equation}
\sum_jW_{mjjl}n_jn_l=-\sum_{ijk}V_{mkij}\Gamma^{(2)}_{\xc,ijkl}.
\end{equation}
 For $W=v$ this is the exchange approximation. Using a static $W$ corresponds to the screened exchange (SEX) approximation, which, as HF, leads to occupation numbers equal to 0 or 1. In order to get fractional occupation numbers we combine this approximation with the power functional to get the screened power functional
\begin{equation}
\sum_jW_{mjjl}n^\alpha_jn^\alpha_l=-\sum_{ijk}V_{mkij}\Gamma^{(2)}_{\text{xc},ijkl},
\label{Eqn:W_x}
\end{equation}
where $W$ has to be considered fixed (which, hence, does not enter into the variational process). The rationale behind this approximation is that the PF will describe strong correlation (or nondynamic correlation, related to the existence of quasi-degenerate states) whereas a static $W$ will describe weak correlation (or dynamic correlation, related to electron screening). Of course double counting problems are possible, as we shall see when discussing the results. In the following we will refer to this approximations as $W$-PF. 
This derivation can be extended also to the COHSEX (Coulomb hole + screened exchange) approximation\cite{PhysRev.139.A796, PhysRevB.34.5390, Hedin_1999, Berger_2021}, which is more commonly used in many-body perturbation theory. This is shown in Appendix \ref{Appendix:A}. The final result is similar to Eq.~(\ref{Eqn:W_x}) with an extra term taking into account the Coulomb hole (COH). 

\section{Models\label{Sec:Models}}
To test the quality of the \wpf functional we use two well-known models in condensed matter physics, namely the one-dimensional Hubbard model and the homogeneous electron gas (HEG). 
\subsection{One-dimensional Hubbard model}
In this work we will consider a Hubbard chain with number of sites $L$ and periodic boundary conditions. The Hamiltonian of the Hubbard model, in second quantization, reads as
\begin{align}
 \hat{H} =& -t \sum_{\langle R,R'\rangle}\sum_{\sigma} \cre{R\sigma} \an{R'\sigma}
 +\frac{U}{2}
 \sum_{R}\sum_{\sigma\sigma^\prime}\cre{R\sigma}\cre{R\sigma^\prime}\an{R\sigma^\prime}\an{R\sigma}.
 \label{Eqn:AppHubbard_HubbardHam}
 \end{align}
Here $\cre{R\sigma}$ and $\an{R\sigma}$ are the creation and annihilation operators for an electron at site $R$ with spin $\sigma$, $U$ is the on-site (spin-independent) interaction, $-t$ is the hopping kinetic energy.
The summation $\sum_{\langle R,R'\rangle}$ is restricted to the nearest-neighbor sites. Due to the translational invariance of the system the natural orbitals have the form
$
\phi_{I\sigma}=\frac{1}{\sqrt{L}}\sum_R e^{\I I R} \varphi_{R\sigma}
$,
where $\varphi_{R\sigma}$ are the site spin-orbitals, 
and the
total energy is a function  of the occupation numbers alone which reads \cite{stefano_thesis}
\begin{align}
E[\{n_{I\sigma}\}]=&\sum_{I} \sum_{\sigma} \epsilon^0_I n_{I\sigma}+
\frac{U}{4} \sum_{IJ}\sum_{\sigma \sigma'} n_{I\sigma} n_{J\sigma'}\nonumber\\
 &-\frac{U}{4} \sum_{IJ}  \sum_{\sigma} n_{I\sigma}^\alpha n_{J\sigma}^\alpha,
 \label{eqn:enFunctHub}
\end{align}
where $\epsilon_I^0=-2t \,\text{cos}\left[2\pi(I-1)/L\right]$ is the non-interacting energy associated with the $I$-th natural orbital,\footnote{Note that this formula is valid only for $L>2$. For $L=2$ we have $\epsilon_1=-t$ and $\epsilon_2=+t$.
}
and we used the PF functional to approximate $\Gamma_{\xc}$, with $0.5\le \alpha\le 1$\cite{PhysRevB.78.201103}. In this work we only considered the spin-symmetric case at one-half filling.

\subsection{HEG}
The HEG Hamiltonian in its spin-explicit form is given by the following expression
\bea
\hat{H}&=&\sum_\sigma\sum_\vect{k} \frac{\vect{k}^2}{2} \hat{c}^\dagger_{\vect{k},\sigma}
\hat{c}_{\vect{k},\sigma}\nonumber\\
&&+\frac{1}{2\Omega}
\sum_{\sigma\sigma'}\sum_{\vect{k}\neq 0,\vect{k}_1,\vect{k}_2} \frac{4\pi}{\vect{k}^2} \hat{c}^\dagger_{\vect{k}_1+\vect{k},\sigma}
\hat{c}^\dagger_{\vect{k}_2-\vect{k},\sigma'}
\hat{c}_{\vect{k}_2,\sigma'}
\hat{c}_{\vect{k}_1,\sigma}+E_b,
\eea
where $\vect{k}$ is a plane wave vector and $\Omega$ is the volume of the unit cell.
Note that to guarantee the charge neutrality of the system a positive background charge has to be included. 
This results in the constant term $E_b$ in the Hamiltonian, which contains the  electron-background interactions.

Due to the translational invariance of the HEG, the
natural orbitals can be chosen to be plane waves. The minimization procedure reduces then to the search for the optimal momentum distribution $n(\vect{k})$, i.e., the occupation number corresponding to the plane-wave natural orbital with wave vector $\vect{k}$. We also note that, due to the rotational invariance, $n(\vect{k})=n(k)$, i.e, the momentum distribution depends only on the magnitude of $\vect{k}$.

The total energy functional per unit volume can be expressed in terms of the momentum distribution as
\be
\frac{E}{\Omega}=\int \frac{d \vect{k}}{(2\pi)^3} \vect{k}^2 n(\vect{k}) -
\int \frac{d\vect{k}d\vect{k}'}{(2\pi)^6}v(\vect{k}-\vect{k}')
f(n(\vect{k}),n(\vect{k}')),
\label{eqn:E_HEG}
\ee
where the first and second terms on the right-hand side are the kinetic energy and the exchange-correlation energy per unit of volume, respectively 
and $v(\vect{k})=4\pi/|\vect{k}|^2$ is the Coulomb potential. Note that the Hartree energy is not included in Eq.~\eqref{eqn:E_HEG} since it is compensated by $E_b$. RDMFT functionals have
already been applied to the homogeneous electron gas \cite{Lathiotakis_PRB07,Csanyi_PRB2000,Csanyi_PRA2002,Pernal_JCP1999}. In particular Lathiotakis \textit{et al.}\cite{PhysRevB.75.195120} studied the performance of the BBC functionals for the  correlation energies and the momentum distribution. 
Within the PF approximation to the 2-RDM we have  $f^{\text{PF}}(n(\vect{k}),n(\vect{k}'))=[n(\vect{k})n(\vect{k}')]^\alpha$ with $0.5\le\alpha\le1$ \cite{PhysRevB.78.201103}, while for the BBC1 functional we have $f^{\text{BBC1}}(n(\vect{k}),n(\vect{k}'))=\sqrt{n(\vect{k})n(\vect{k}')}[1 - 2 \theta(|\vect{k}| - k_F) \theta(|\vect{k}'| - k_F)]$, where $\theta$ is the Heaviside step function. Within the screened power functional approximation, instead, 
$f^{\text{\wpf}}(n(\vect{k}),n(\vect{k}'))=\bar{W}(\vect{k}-\vect{k}')/v(\vect{k}-\vect{k}')[n(\vect{k})n(\vect{k}')]^\alpha$,
i.e., 
the Coulomb potential $v(\vect{k})$ in the PF is replaced by the screened interaction $\bar{W}(\vect{k})$.
In the following we will assume $\bar{W}$ to be the static limit of the dynamically screened interaction $W$ given by
 \be
\bar{W}(\vect{k})=\frac{v(\vect{k})}{1-v(\vect{k})P^0(0,\vect{k})},
\ee
with the static RPA polarizability given by the Lindhard formula\cite{osti_4405425}
\be
P^0(0,\vect{k})=\frac{2k_F}{4\pi^2}\left\{-1+\frac{k_F}{2k}\left(1-\frac{k^2}{4k_F^2}\right)\ln \left[\frac{1-k/(2k_F)}{1+k/(2k_F)}\right]^2
\right\},
\ee
where $k_F$ is the Fermi momentum given by $k_F=(9\pi /4)^{1/3}/r_s$ and $r_s$ the Wigner radius. 

For the HEG, the matrices given in Eqs.~\eqref{eqn:lambdaR} and \eqref{Eqn:Lagrangian_A} are diagonal in the basis of natural orbitals and their diagonal elements are the EKT removal and addition energies
respectively given by
\be
\epsilon^R(\vect{k})=\frac{k^2}{2}-\frac{1}{n(\vect{k})}\int \frac{d \vect{k}'}{(2\pi)^3}v(\vect{k}-
\vect{k}') f(n(\vect{k}),n(\vect{k}'))
\label{Eqn:VQP1}
\ee
and
\bea
\epsilon^A(\vect{k})&=&\frac{k^2}{2}-\frac{1}{1-n(\vect{k})}\int \frac{d \vect{k}'}{(2\pi)^3}v(\vect{k}-
\vect{k}') n(\vect{k}')\nonumber\\
&&+\frac{1}{1-n(\vect{k})}\int \frac{d \vect{k}'}{(2\pi)^3}v(\vect{k}-
\vect{k}') f(n(\vect{k}),n(\vect{k}')),
\label{Eqn:CQP1}
\eea
where we used the approximation in Eq.~\eqref{eqn:2rdmJK} for the 2-RDM.

\section{Computational details\label{Sec:Computation}}
For the Hubbard model with a finite number of sites we use the Lanczos method\cite{re:alvarez08} for the calculation of the exact one-body Green's function, from which we get all the quantities of interest for this work. 
This poses a limit to the number of sites we can treat, which in our case is $L=12$. 
For the infinite chain ($L\rightarrow\infty$) we use the Bethe ansatz.\cite{PhysRevLett.20.1445} For the total energy minimization within the approximate Power functional we use the direct minimization for finite sites using the MATHEMATICA package \cite{Mathematica}. In the case of the infinite chain instead we use the same strategy used for the HEG, which we describe in the following.

For the HEG, the functional to be minimized can be written as\cite{Lathiotakis_PRB07}
\bea
\frac{\mathcal{F}}{\Omega}&=&\int \frac{d \vect{k}}{(2\pi)^3} (\vect{k}^2-2\mu)n(\vect{k})\nonumber\\
&&-
\int \frac{d\vect{k}d\vect{k}'}{(2\pi)^6}v(\vect{k}-\vect{k}')
f(n(\vect{k}),n(\vect{k}'))+\mu,
\label{eqn:funcE}
\eea
where $\mu$ is the Lagrange multiplier which enforces the condition $\sum_in_i=N$. 
From the stationarity condition
\bea
\frac{\delta(\mathcal{F}/\Omega)}{\delta n(\vect{k})}&=& \frac{1}{(2\pi)^3} (\vect{k}^2-2\mu)\nonumber\\
&&-2
\int \frac{d\vect{k}'}{(2\pi)^6}v(\vect{k}-\vect{k}')
\partial_x f(x,n(\vect{k}'))\big|_{x=n(\vect{k})}\nonumber\\
&=&0
\label{Eqn:functionalderivative}
\eea
and using the PF approximation (i.e., $f(n(\vect{k}),n(\vect{k}'))=[n(\vect{k})n(\vect{k}')]^\alpha$) we can obtain the following integral equation\footnote{Note that for $\alpha=1$ Eq.~\eqref{Eqn:functionalderivative} cannot be used to derived the fixed-point equation \eqref{eqn:fixeP}. In this case the solution is given by the HF solution $n(\vect{k})=\theta(k_F-|\vect{k}|)$.}
\be
n(\vect{k})
=\left\{\frac{\int d\vect{k}'/(2\pi)^3 v(\vect{k}-\vect{k}')
 [n(\vect{k}')]^\alpha} {\vect{k}^2/2-\mu}\right\}^{\frac{1}{1-\alpha}}.
 \label{eqn:fixeP}
\ee

Similar equations can be obtained using the BBC1 and \wpf functionals.
The minimization of the energy functional is thus transformed into a fixed-point problem that can be solved iteratively starting from a reasonable guess for $n(\vect{k})$ (e.g., the non-interacting distribution). The Lagrange multiplier $\mu$ is determined through an iterative procedure by requiring that the momentum-distribution function integrates to the correct number of electrons.
The condition $0\le n_i \le 1$ is enforced at each step.
The integral in Eq.~\eqref{eqn:fixeP} and the evaluation of Eqs.~\eqref{Eqn:VQP1} and \eqref{Eqn:CQP1} are performed numerically using the MATHEMATICA package \cite{Mathematica}.

\begin{figure}
\centering
   \includegraphics[width=0.95\columnwidth]{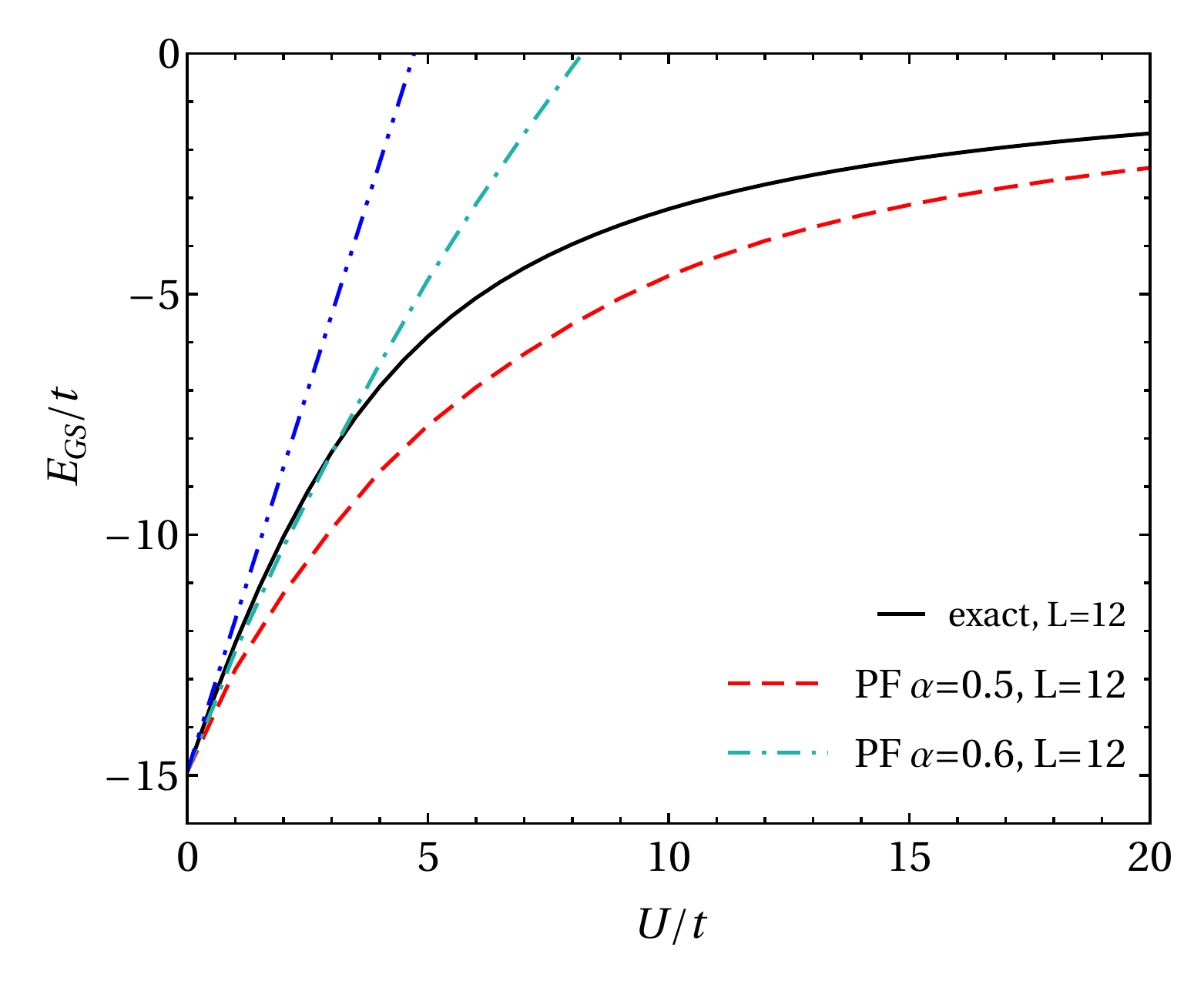} 
   \vspace{-10pt}
\caption{Total energy as function of $U/t$ for a 12-site Hubbard chain. Exact vs PF ($\alpha=0.5$) and PF ($\alpha=0.65$).}
\label{fig:hubbarden}
\end{figure}

\begin{figure}[t]
\centering
   \includegraphics[width=0.95\columnwidth]{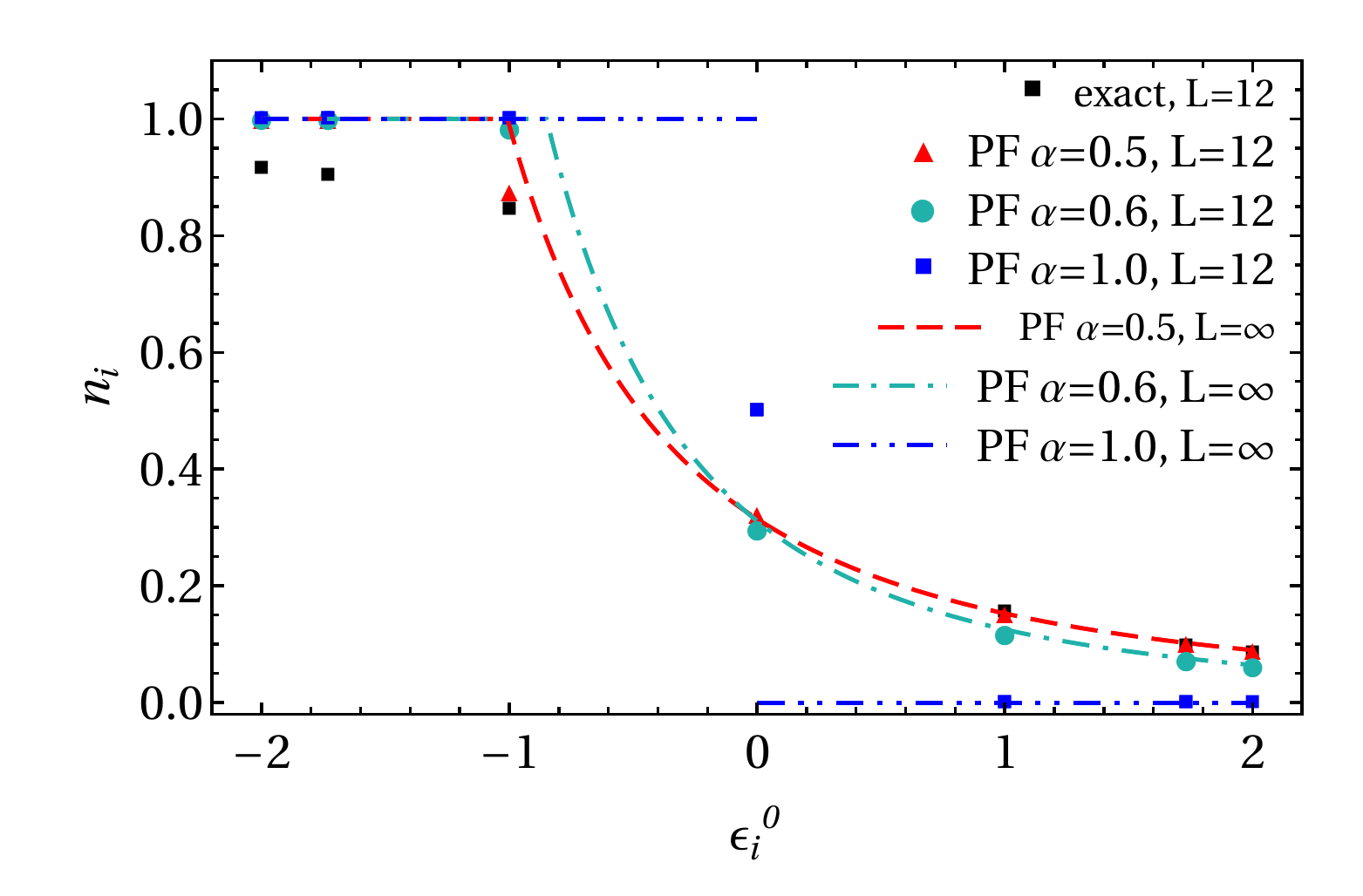}
   \includegraphics[width=0.95\columnwidth]{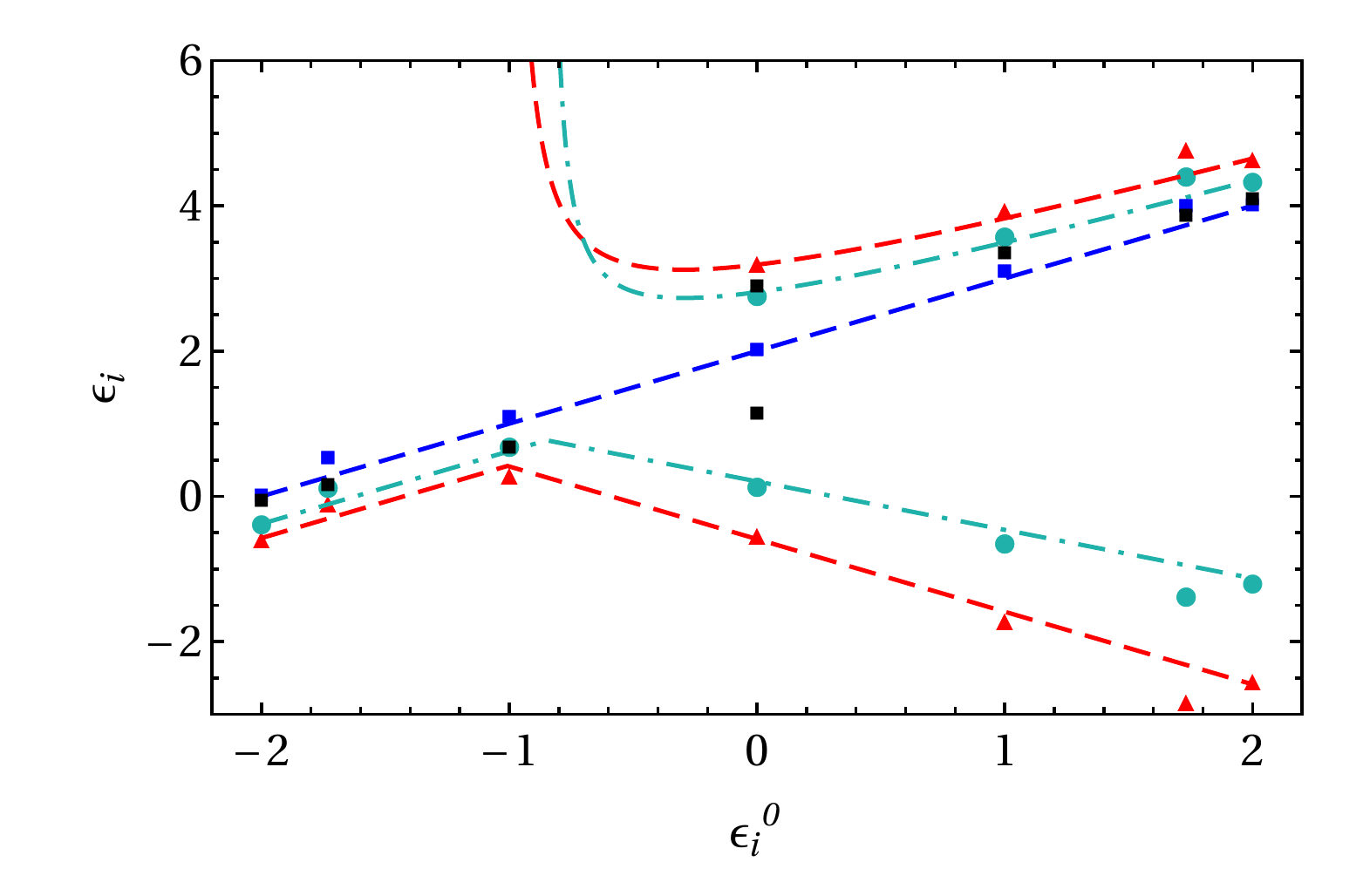}
 \caption{Occupation numbers (top panel) and removal/addition energies $\epsilon_i$ (bottom panel)  for a 12-site Hubbard chain and the infinite Hubbard chain at $U/t=4$. Exact results are reported with black squares for the 12-site Hubbard chain.}
\label{fig:hubbardocc}
\end{figure}
\begin{figure}[t]
   \includegraphics[width=0.95\columnwidth]{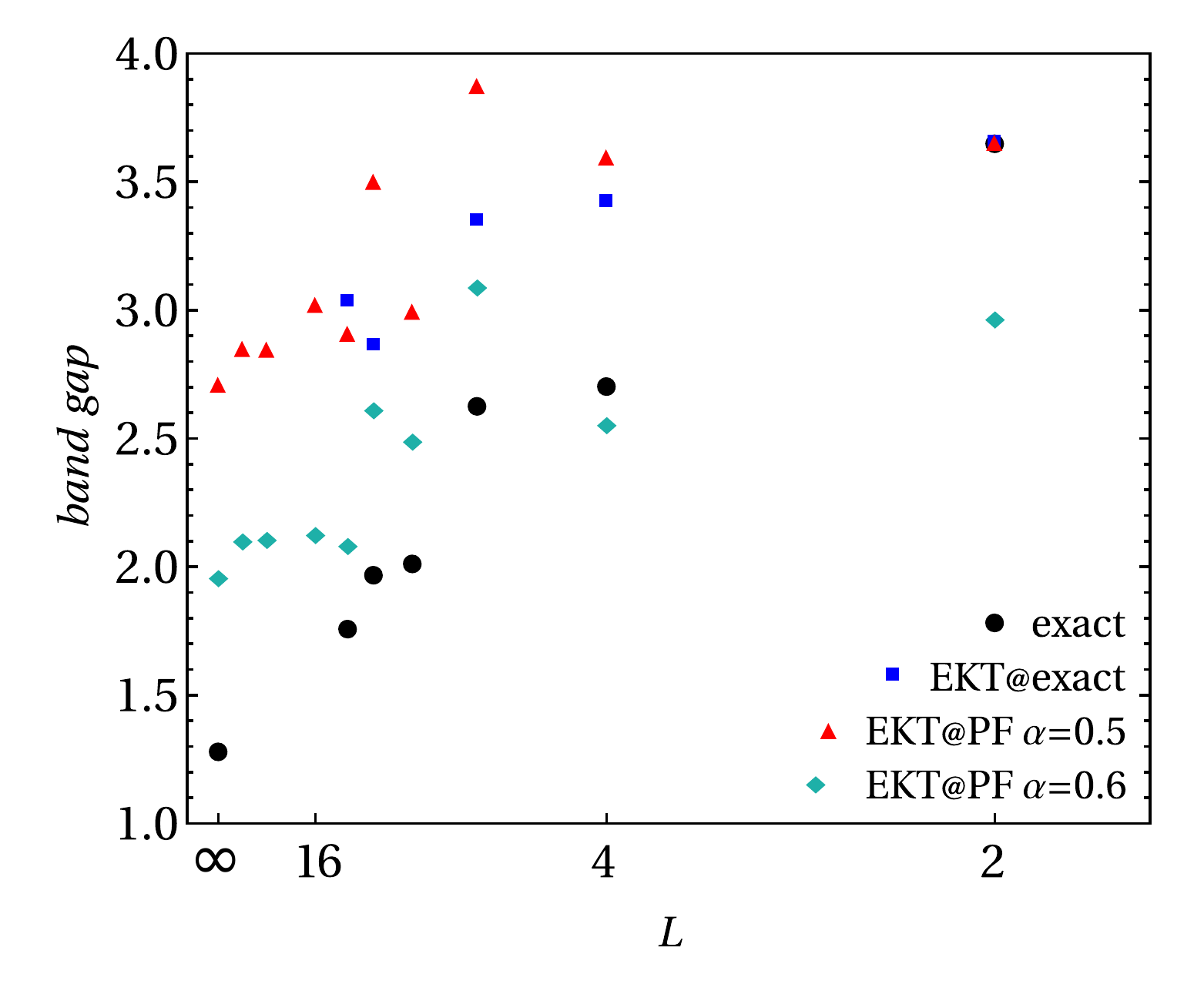}  
   \vspace{-10pt}
\caption{Fundamental band gap of the one-dimensional Hubbard chain as function of the length $L$ for $U/t=4$: exact results, EKT@exact, and EKT@PF for both $\alpha=0.5$ and $0.6$.}
\label{fig:hubbard}
\end{figure}

\section{Results and discussion\label{Sec:Results}}
In this section we will assess the quality of the PF and $W$-PF functionals by analyzing the total energies, natural occupation numbers, band gaps/band widths  from the EKT using the Hubbard model and the HEG. 
\subsection{Hubbard model}
In Fig.~\ref{fig:hubbarden} we report the total energy of a 12-site Hubbard chain, obtained from the direct minimization
of the total energy functional given in Eq.~\eqref{eqn:enFunctHub}, as function of $U/t$. We notice that only the PF with $\alpha=0.5$ gives the correct limit for the large interaction limit $U/t\rightarrow\infty$,\footnote{For $U/t\rightarrow\infty$ the exact energy vanishes, in fact, each electron localize on one site and double-occupancy are not allowed (we assumed that the site orbital energy is zero). 
In the approximate functional given in Eq.~\eqref{eqn:enFunctHub}, for $U/t\rightarrow\infty$ the kinetic term is negligibly small compared to the exchange-correlation term. Since the $n_i$ are symmetric, the optimal occupation numbers are $n_i=N/L=1/2$. Substituting this values in the energy functional we obtain $E=0$.} while for larger values of $\alpha$ the result diverges. The value $\alpha=0.5$ also gives the best ``global" result. The total energy is hence quite sensitive to the value of $\alpha$. This is a general trend that is independent of the number of sites.

The occupation numbers which minimize the total energy functional are reported in Fig.~\ref{fig:hubbardocc} for a 12-site Hubbard chain at $U/t=4$. Their trends closely resemble those of the infinite chain, also reported in the figure. The PF gives for some states pinned occupation numbers, i.e., $n_i=1$. In general there is not a large difference between the results obtained with $\alpha=0.5$ and $\alpha=0.6$, except for the occupation of the top valence orbital (at $\epsilon_i=-1$). However we notice a significant difference in the EKT band gap. This is shown in Fig.~\ref{fig:hubbard}, where we present the band gap of the one-dimensional Hubbard model as function of the number of sites $L$. Exact results are compared with those obtained from the EKT using exact RDMs (EKT@exact) and RDMs obtained from the PF approximation (EKT@PF). For the Hubbard dimer, exact, EKT@exact, EKT@PF($\alpha=0.5$) give the same band gap.
For more than two sites ($L>2$) EKT@PF($\alpha=0.5$) shows the same trend as EKT@exact, i.e. a systematic overestimation of the exact band gap. 
The PF($\alpha=0.6$), instead, gives results closer to the exact ones, which points to an error cancellation between the approximate nature of the EKT equations and the approximation to the 1- and 2-RDMs. 
Introducing screening will decrease the gap. For example, for the Hubbard dimer screening in the PF($\alpha=0.5$) has a similar effect than using $\alpha=0.6$ in the PF. 

The BBC1 functional produces results (not reported in Fig.~\ref{fig:hubbard}) in between the results obtained with PF($\alpha=0.5$) and PF($\alpha=0.6$). 
Of course one should be careful  to extrapolate these findings to real materials. This model, indeed, is peculiar because the Power functional, which contracts the four-point two-body density matrix to two points only, i.e., $\Gamma^{(2)}_{\xc,ijkl} = -n^\alpha_in^\alpha_j\delta_{ik}\delta_{jl}$,
is a good approximation due to the topology of the system; moreover the basis of natural orbitals is also the basis which diagonalizes the $\mathbf{\Lambda}^{R/A}$ matrices. These features are not generally true in a real system. However, the fact that the EKT method overestimates the exact band gap seems a general feature, as pointed out in Ref.~\onlinecite{frontiers_2021}, and an important finding. Of course this also questions its applicability to metals, where there is no band gap. We shall investigate this point with the example of the HEG in the next subsection.

\begin{figure}[t]
\centering
         \includegraphics[width=0.95\columnwidth]{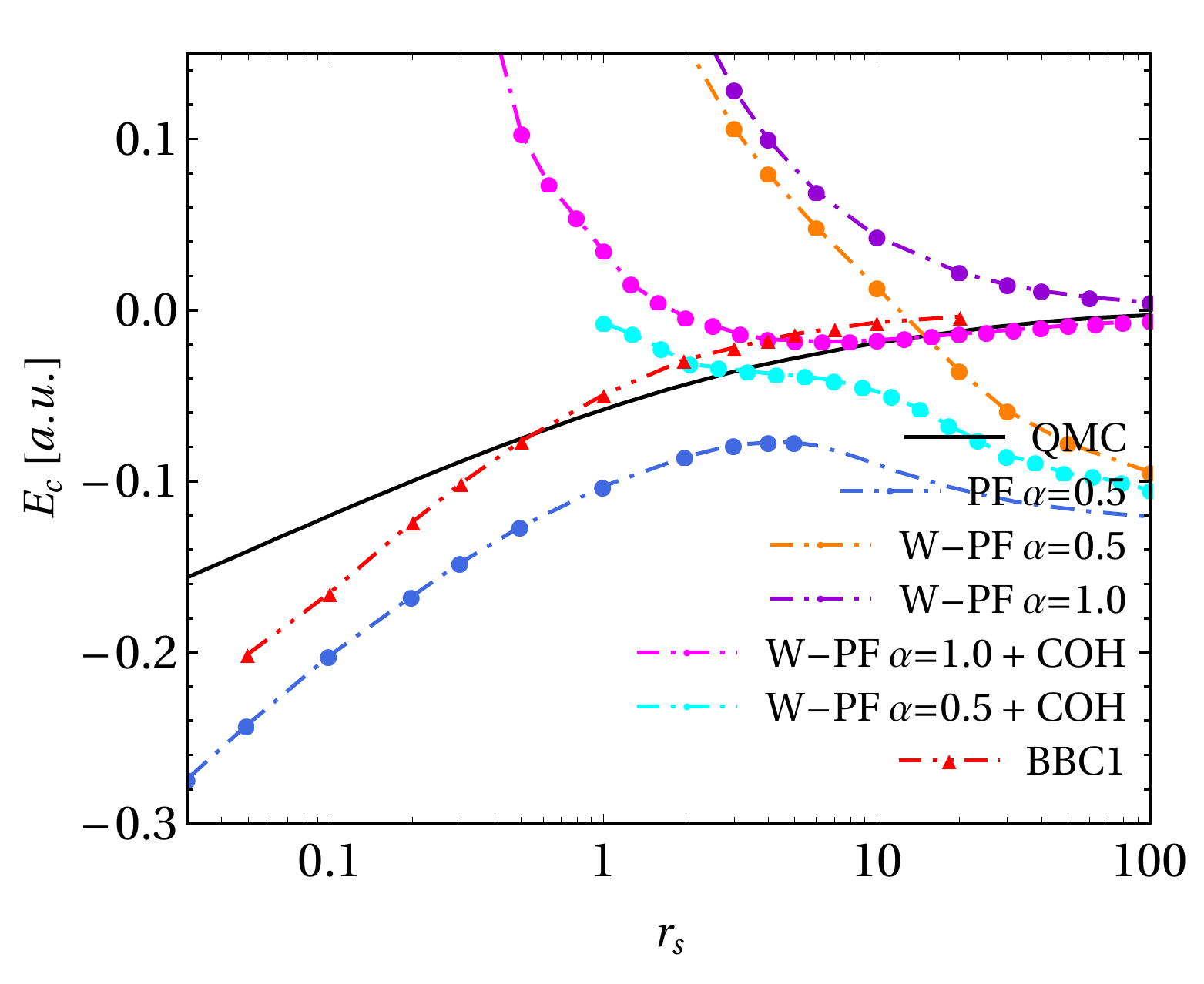}
      \vspace{-10pt}
\caption{Correlation energy of the HEG as a
function of $r_s$ calculated with the PF and \wpf functionals. The Coulomb hole correction to the \wpf is also reported (W-PF+COH). 
The quantum Monte Carlo result (QMC) corresponds to the Perdew-Wang fit \cite{PhysRevB.45.13244} of the DMC data of Ortiz and
Ballone \cite{PhysRevB.50.1391,PhysRevB.56.9970}.
The blue dashed-dotted line, for $r_s<5.77$, is obtained numerically and corresponds to the
results by Csányi and Arias\cite{PhysRevB.61.7348} employing the M\"{u}ller functional. The continuation for $r_s>5.77$, is the analytical results of Cioslowski and Pernal\cite{doi:10.1063/1.479623}.
}
\label{heg}
\end{figure}
\begin{figure}[t]
\centering
   \includegraphics[width=0.95\columnwidth]{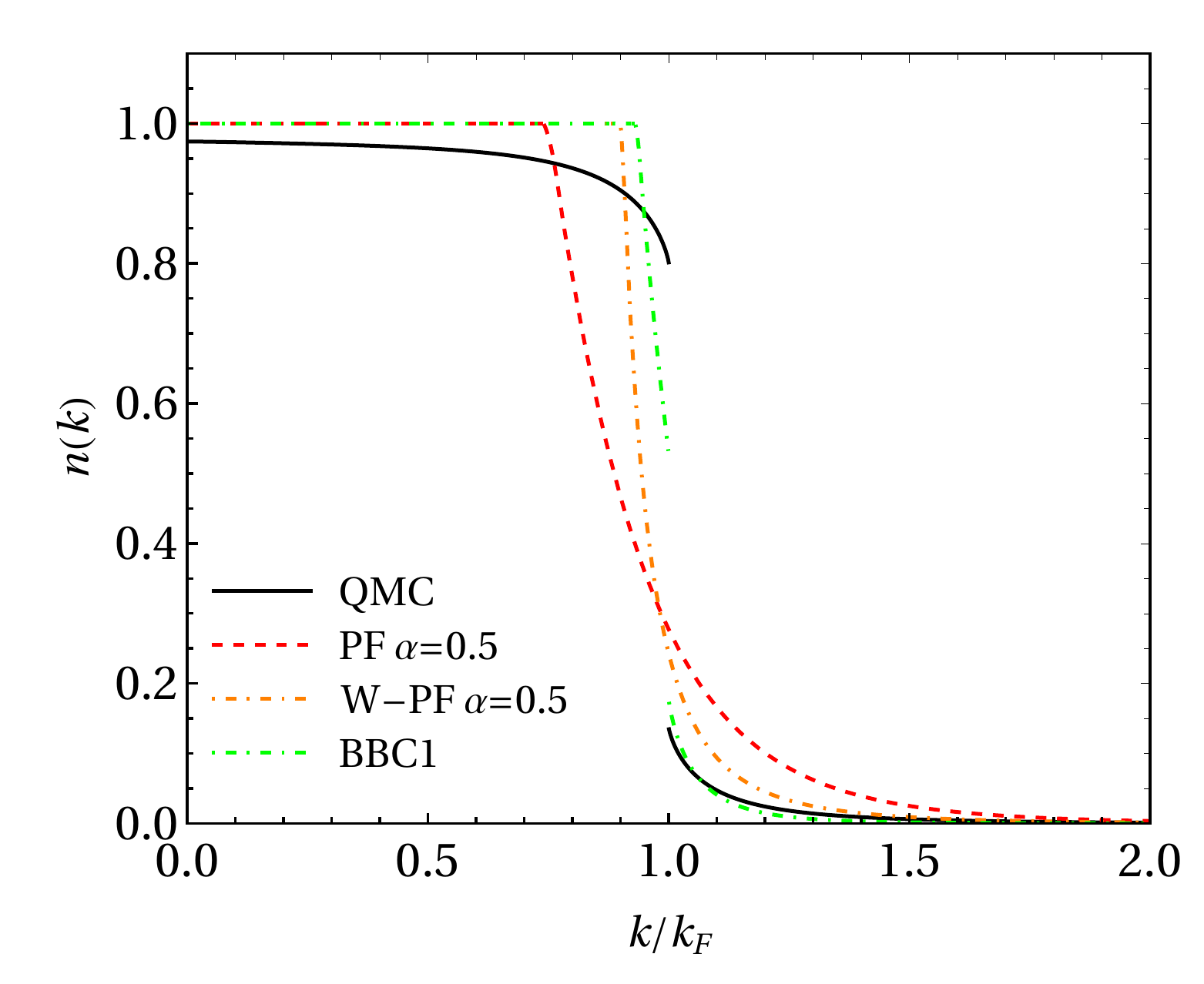}
   \vspace{-10pt}
\caption{Momentum distribution of the HEG for $r_s=3$: QMC vs PF and \wpf for $\alpha=0.5$ and BBC1. The QMC momentum distribution is taken from Ref.~\onlinecite{PhysRevB.66.235116}. 
}
\label{fig:heg_nk}
\end{figure}

\begin{figure*}
\centering
   \includegraphics[width=0.47\textwidth]{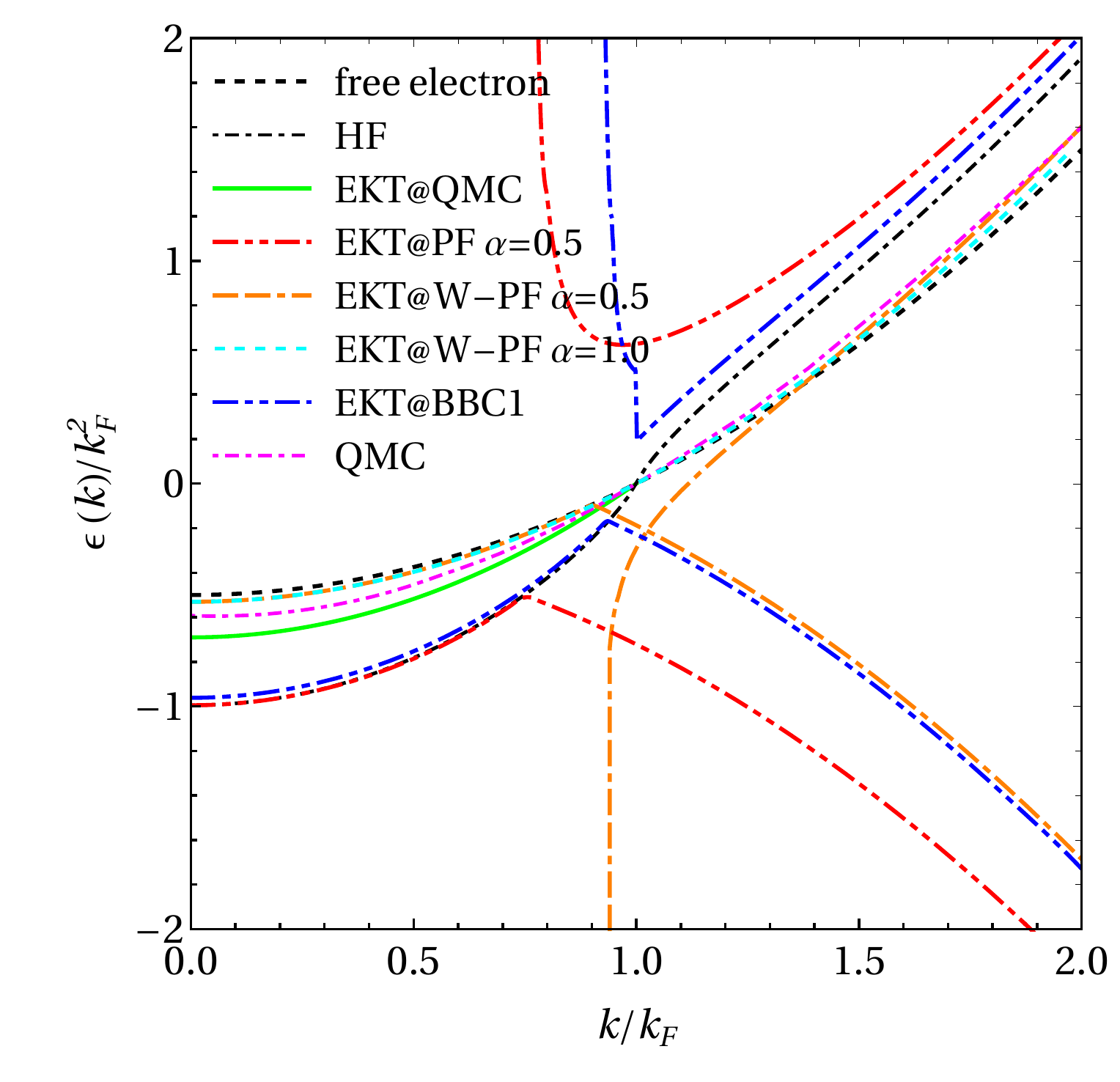}
   \includegraphics[width=0.47\textwidth]{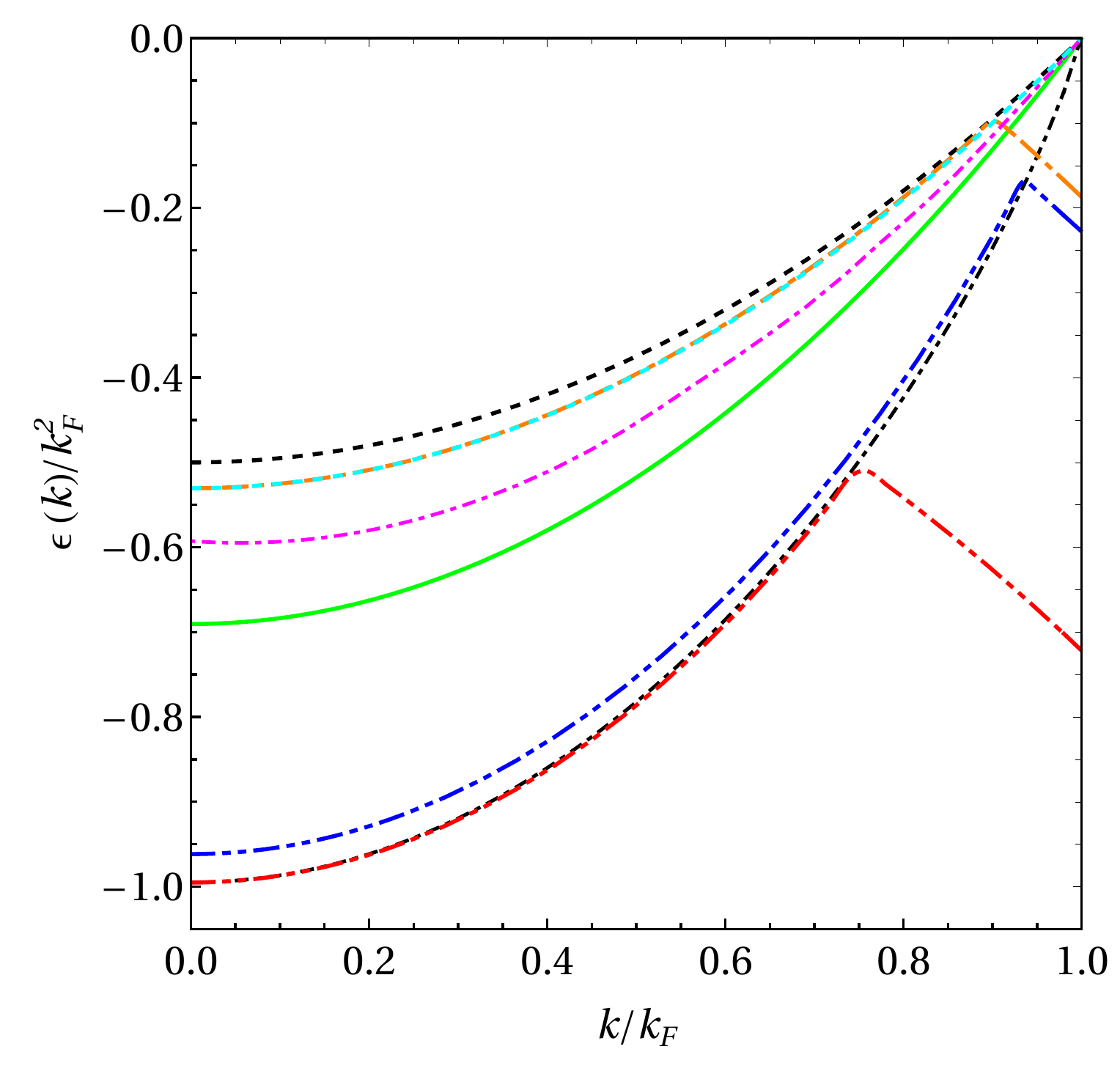}\\
   \includegraphics[width=0.33\textwidth]{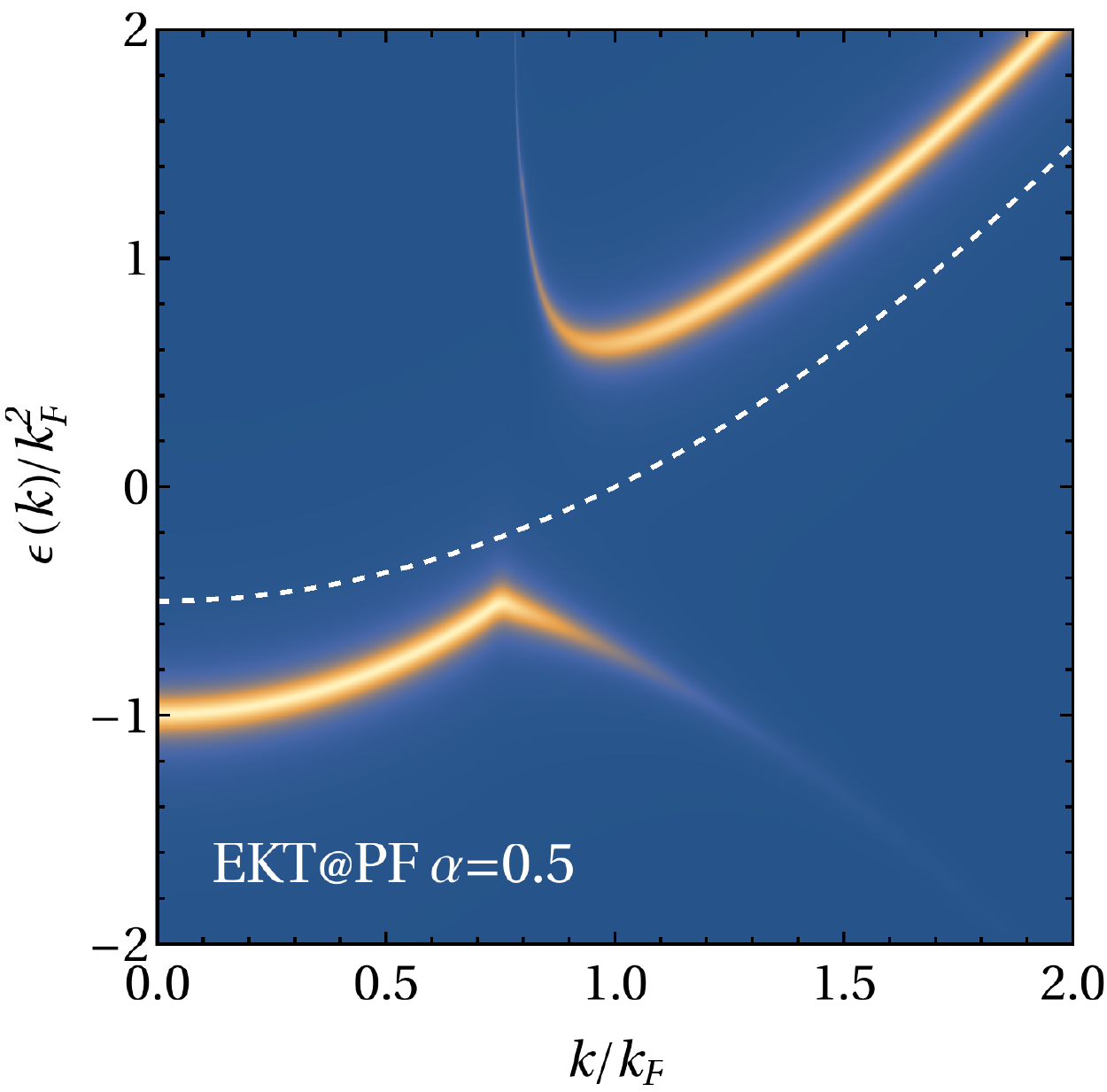}
   \includegraphics[width=0.33\textwidth]{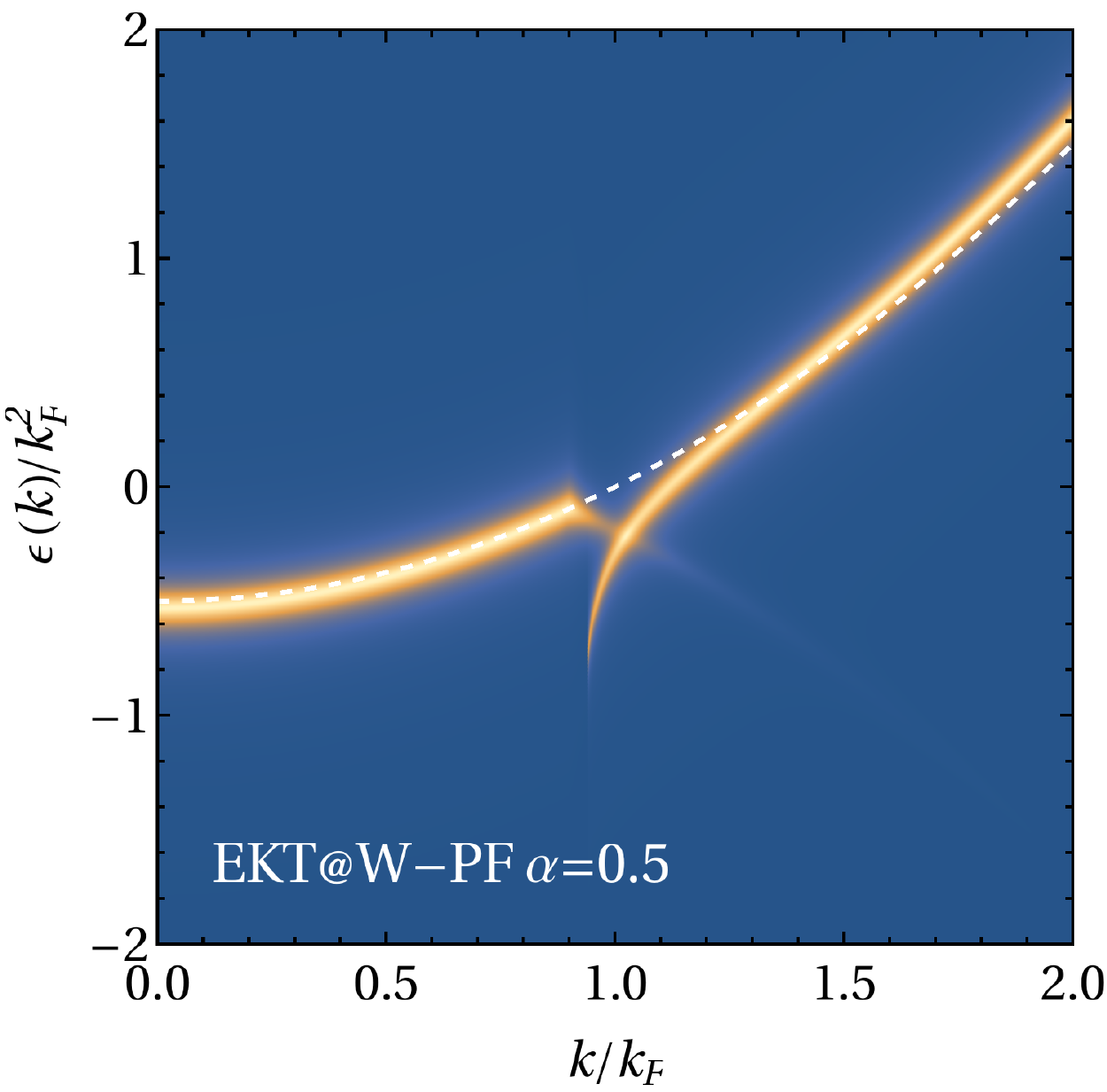}
   \includegraphics[width=0.33\textwidth]{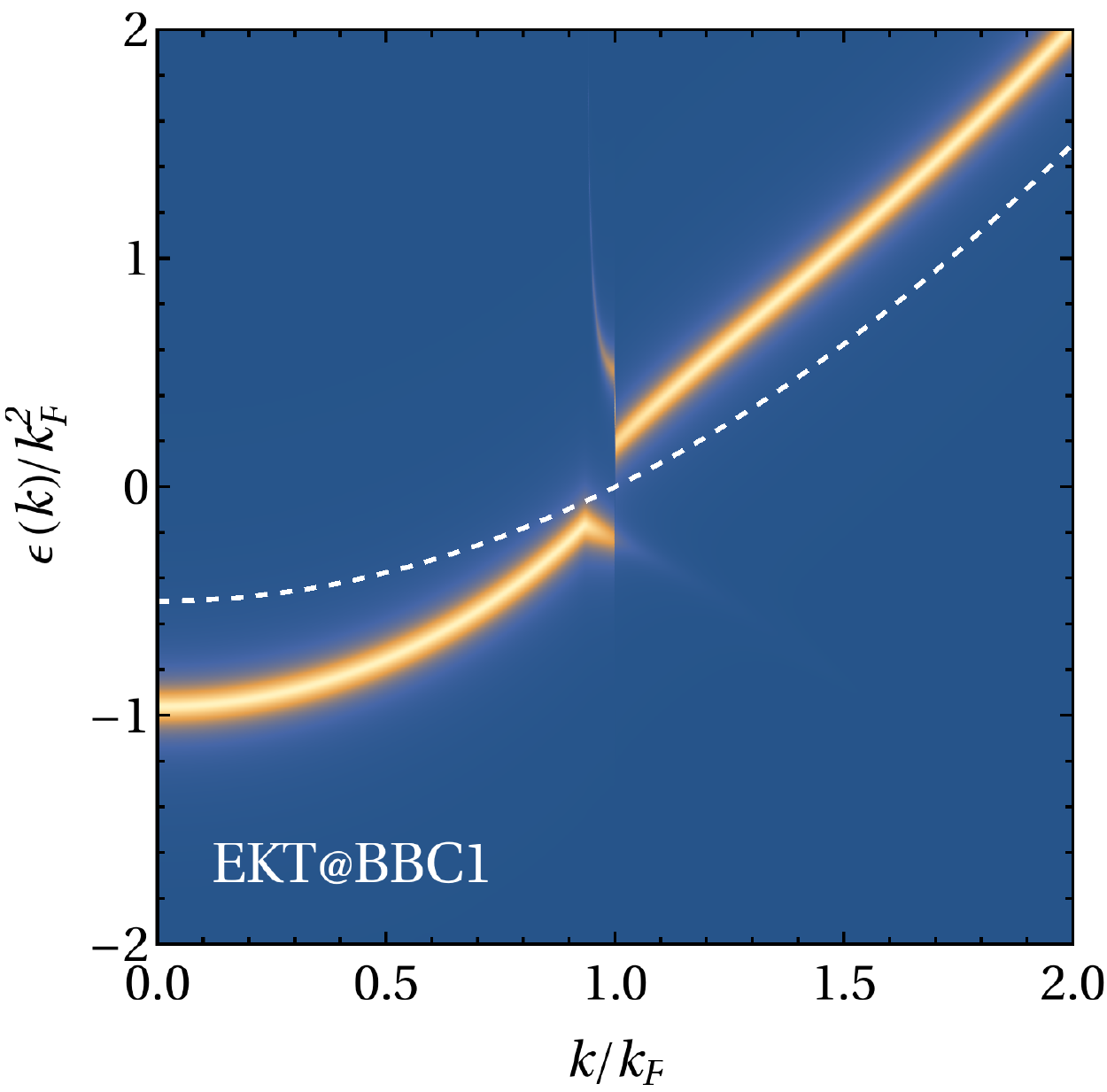}
\caption{Quasiparticle dispersion $\epsilon(k)/k_F^2$ for $r_s=3$: EKT@\wpf, EKT@PF and EKT@BBC1 are compared with EKT@QMC results extracted from Ref~\onlinecite{PhysRevB.90.035125} and QMC quasiparticle dispersion from Ref.~\onlinecite{PhysRevLett.127.086401}. The free electron and Hartree-Fock dispersions are also reported.
In the bottom panels we report the momentum resolved spectral function $A(\vect{k},\omega)$. The free-electron dispersion is indicated with a dashed white line
}
\label{fig:hegDisp2}
\end{figure*}

\subsection{Homogeneous electron gas\label{HEG}}
We first examine the correlation energy reported in Fig~(\ref{heg}). Our reference is the Monte Carlo (QMC) correlation energy from Ref.~\onlinecite{PhysRevB.45.13244}. We compare the results obtained using PF and $W$-PF. 
In Ref.~\onlinecite{PhysRevA.79.040501} it is shown that the correlation energy of the HEG is well reproduced by the PF with values of $\alpha$ between 0.55 and 0.58 depending on the value of the Wigner-Seitz radius $r_s$. 

For $\alpha=1$ the \wpf corresponds to the screened exchange approximation (SEX) and it gives positive correlation energies for all densities. Considering $\alpha<1$ decreases the correlation energy.
Moreover, we note that for $\alpha=0.5$ this functional inherits the incorrect high density limit ($r_s\rightarrow 0$) of the SEX and the incorrect low density limit  ($r_s\rightarrow\infty$) of the PF($\alpha=0.5$).  A similar scenario is observed using the Coulomb hole correction in the $W$-PF (see App.~\ref{Appendix:A}).
The correlation energy obtained using the BBC1, also reported in Fig~(\ref{heg}), instead, performs very well over a wide range of $r_s$ values.
 In Fig.~\ref{fig:heg_nk} we report the momentum distribution $n(\vect{k})$ for $r_s=3$ calculated within QMC\cite{PhysRevB.66.235116} (our reference), the PF and the $W$-PF functional (both with $\alpha=0.5$). The PF functional is not able to describe the characteristic
discontinuity  of the exact momentum distribution at the Fermi momentum $k_F$. This is a general feature of the M\"{u}ller-like functionals. In Ref.~\onlinecite{doi:10.1063/1.479623}, for example, it was shown that for values of $r_s<5.77$ the M\"{u}ller functional produces occupation numbers pinned to 1 for values of $k$ smaller than a characteristic value $k_p$. For $k>k_p$ the occupation decreases monotonically to zero without discontinuity. As pointed out in Ref.~\onlinecite{PhysRevB.75.195120} only the BBC functionals have been reported to reproduce this feature. This is indeed what we find by using the BBC1, also reported in Fig.~\ref{fig:heg_nk}. The \wpf improves the situation in the sense that it enlarges the range $0<k<k_p$, but it cannot reproduce the discontinuity either.

\begin{figure*}
\centering
   \includegraphics[width=0.47\textwidth]{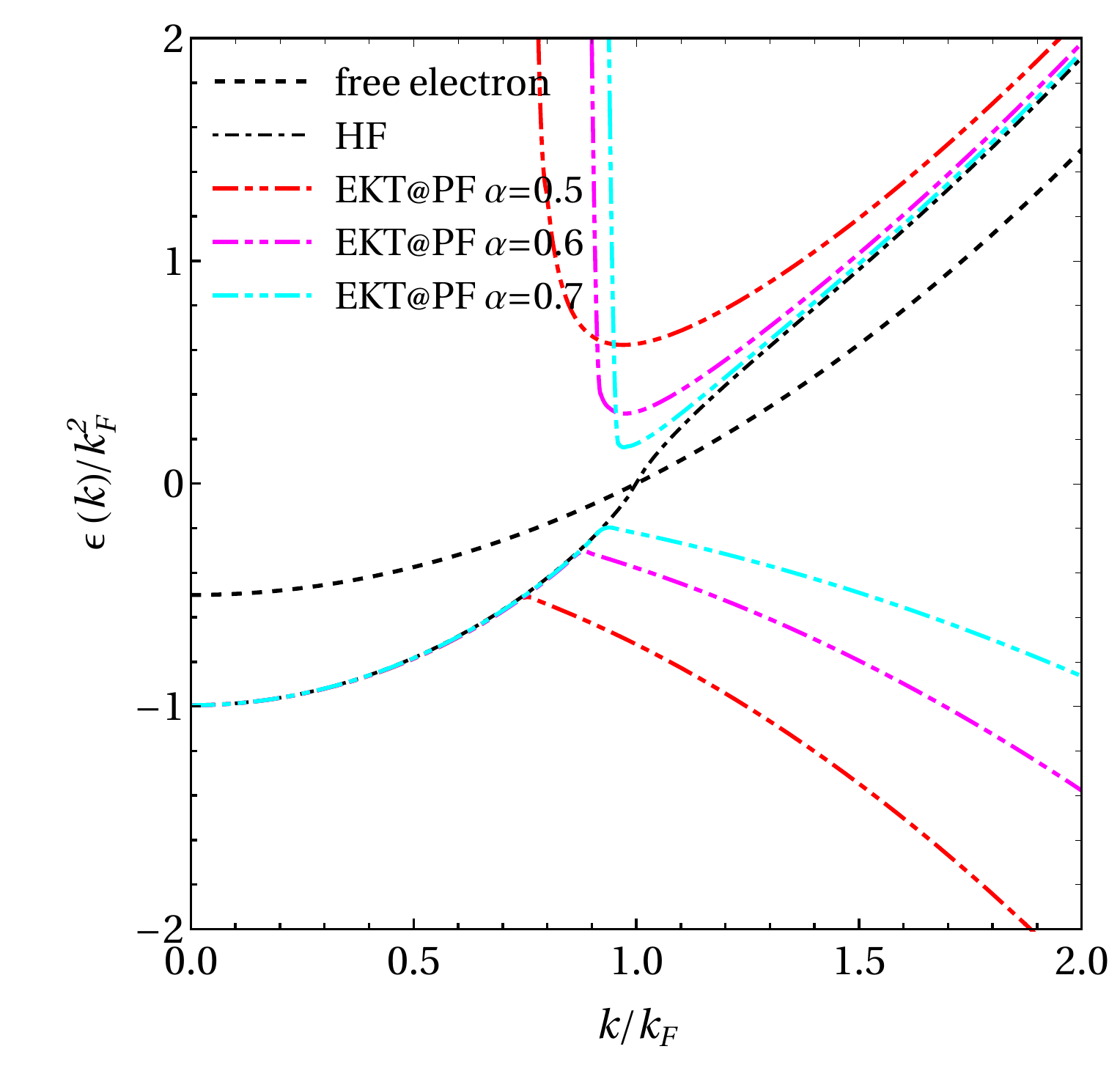}
   \includegraphics[width=0.47\textwidth]{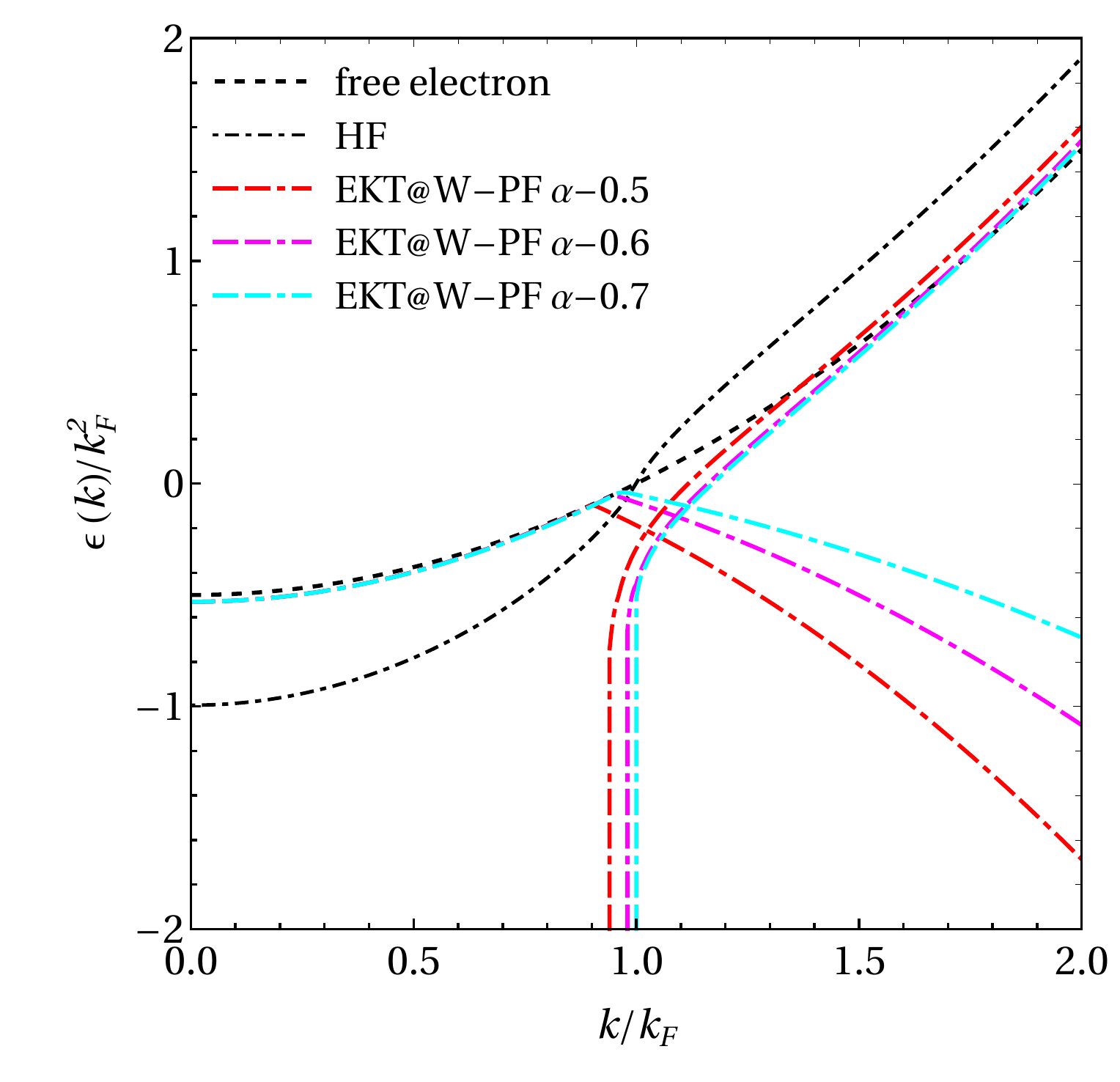}
   \vspace{-10pt}
\caption{Dependence of the EKT@PF and EKT@\wpf QP energy dispersion on the parameter $\alpha$.}
\label{fig:varA}
\end{figure*}

In Fig.~\ref{fig:hegDisp2} we report the QP dispersion curve obtained with the EKT.
The first remarkable feature that we observe is the opening of an unphysical band gap.
The EKT@PF QP dispersion is very close to the HF one for $k<k_p$. 
Due to the fact that the PF is not able to well reproduce the momentum distribution near $k=k_F$ also the QP dispersion is strongly deformed near the Fermi momentum ($k_p<k<k_F$). The range of deformation is instead smaller for the \wpf. 

\begin{figure}
\centering
   \includegraphics[width=0.47\textwidth]{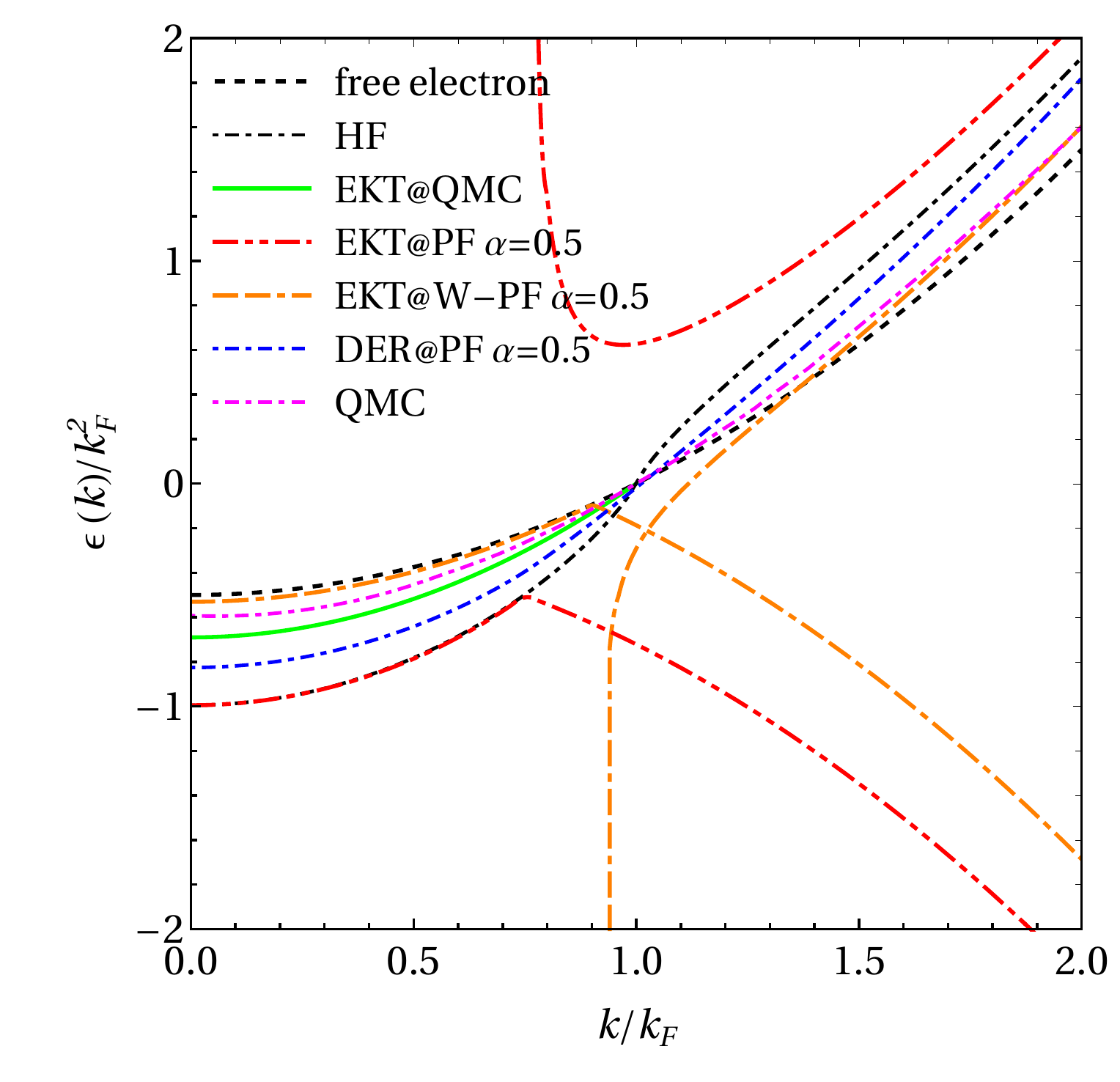}
   \vspace{-10pt}
\caption{Quasiparticle dispersion $\epsilon(k)/k_F^2$ for $r_s=3$: EKT@\wpf and EKT@PF are compared with the DER@PF.}
\label{fig:hegDispDER}
\end{figure}
Moreover, introducing screening in the PF functional reduces the overestimation of the band width, which becomes smaller than the EKT@QMC result. This finding points to an overscreening in the \wpf. Indeed introducing a parameter that reduces the screening in the \wpf would bring the results in line with the EKT@QMC results, as we show in App.~\ref{Appendix:B}. Nevertheless, the \wpf correctly closes the band gap in the HEG; it would hence be interesting to apply the EKT@\wpf to realistic systems and in particular to gapped materials. This study is currently in progress. 

Interestingly the EKT band dispersion obtained using the BBC1 functional is rather bad, at least for the valence part, with respect to the QMC results. Although, similarly to the \wpf, the BBC1 decreases (but does not close completely) the artificial band gap that the exact EKT method opens in the HEG, the BBC1 band dispersion is quite different from the \wpf dispersion. This is the case also for other values of $r_s$. 
We compare our results with the QMC quasiparticle energies obtained from Ref.~\onlinecite{PhysRevLett.127.086401}: \wpf shows a similar dispersion curvature as QMC, whereas BBC1 is very similar to HF.
We note that also increasing $\alpha$ in the PF functional tends to close the band gap, with $\alpha=1$ showing no gap (see Fig.~\ref{fig:varA}). However this is the HF solution which is not a good approximation to the 1- and 2-RDM.  

In Fig.~\ref{fig:hegDispDER} we also compare our results with the method proposed in Ref.~\onlinecite{Sharma13} (referred to as the DER method). For valence QP energies this method gives the same expression as the EKT@\wpf with $W$ replaced by the parameter $\alpha$ of the PF functional. The results of the DER method, however, are closer to those obtained using EKT@PF than to the ones obtained using EKT@\wpf, showing that screening has a stronger impact than the $\alpha$ parameter. It would be interesting to explore the use of a static value of $W$ to fix the parameter $\alpha$ in an \textit{ab-initio} manner, but this is beyond the scope of the present work.

\section{Conclusions and Perspectives\label{Sec:Conclusions}}
In this work we explored the influence of the  approximations to the 1-RDM and 2-RDM on the removal/addition energies calculated using the Extended Koopmans' Theorem within reduced density matrix functional theory. In particular we have focused on the power functional (PF) approximation to the 2-RDM proposed by Sharma and coworkers, which is often employed in solids. Using the one-dimensional Hubbard chain and the HEG as test systems, we explored the sensitivity of the results to the $\alpha$ parameter of this approximation and the impact of introducing screening in the PF (\wpf).  In particular we found that: i) in the Hubbard chain the parameter $\alpha=0.5$ is the best choice for any number of sites when looking at the total energy, the natural occupation numbers, and the EKT band gap; ii) the EKT energies obtained using exact density matrices and PF($\alpha=0.5$) density matrices show a systematic overestimation of the band gap, whereas increasing $\alpha$ yields  a better agreement with the exact band gap, pointing to an error cancellation; iii) introducing screening reduces the gap in the Hubbard model, and improves the quasiparticle dispersion and the band width in the HEG. Although the \wpf does not have a rigorous foundation, our results point to some interesting features for the description of quasiparticle energies. We have also explored the performances of the BBC1 functional, which, as already reported in literature, well reproduces the correlation energy of the HEG over a wide range of $r_s$ and, at the same time, shows a discontinuity in the occupation number distribution at the Fermi level, as in the exact case. 
The trend of the EKT removal and addition energies obtained using the BBC1 is similar to the one observed using the \wpf for the HEG, in particular the fact that the band gap tends to close compared to PF. Nevertheless the band dispersion and band width are quite poor and very similar to the one calculated using HF, contrary to \wpf which performs quite well.  It would be worthwhile to explore the performance of these two functionals on realistic systems. This work is currently in progress.

\begin{acknowledgments}
This study has been supported through the EUR grant NanoX ANR-17-EURE-0009 in the framework of the ``Programme des Investissements d'Avenir" and by ANR (project ANR-18-CE30-0025 and ANR-19-CE30-0011).
\end{acknowledgments}

\appendix
\section{The Coulomb hole plus screened-exchange PF\label{Appendix:A}}
Let us consider the correlation part of $W$, i.e., $W_p=W-v$. Within the COHSEX approximation to the self-energy the correlation contribution of Eq.~\eqref{Eqn:general_G2}, in the limit  $t_4=t_1^{+++}$, reads
\begin{widetext}
\begin{eqnarray}
\sum_{i}\int dt_2\Sigma_{\text{c},mi}(t_1t_2)G_{il}(t_2-t_4)&=&\frac{\I}{2}\sum_{ijk}\int dt_2 G_{kj}(t_1-t_2)W_{p,mjki}(\omega=0)\left[\delta(t_1-t_2+\eta)+\delta(t_1-t_2-\eta)\right]G_{il}(t_2-t_4)\nonumber\\
&=&\frac{\I}{2}\sum_{ijk}\left[G_{kj}(-\eta)W_{p,mjki}G_{il}(t_1+\eta-t_4)+G_{kj}(\eta)W_{p,mjki}G_{il}(t_1-\eta-t_4)\right]\nonumber\\
&=&\frac{\I}{2}\sum_{ijk}\left[G_{kj}(-\eta)W_{p,mjki}G_{il}(-\eta)+G_{kj}(\eta)W_{p,mjki}G_{il}(-\eta)\right]\nonumber\\
&=&\frac{\I}{2}\sum_{ijk}\left[G_{kj}(-\eta)W_{p,mjki}G_{il}(-\eta)+G_{kj}(-\eta)W_{p,mjki}G_{il}(-\eta)-\I \delta_{kj}W_{p,mjki}G_{il}(-\eta)\right]\nonumber\\
&=&\frac{\I}{2}\sum_{ijk}\left[-2 W_{p,mjki}\gamma_{kj}\gamma_{il}+ \delta_{kj}W_{p,mjki}\gamma_{il}\right]
\label{Eqn:general_G2_2}
\end{eqnarray}
\end{widetext}
By adding the exchange contribution to $\Gamma^{(2)}$ and using the basis of natural orbitals, we arrive at 
\begin{align}
&-\sum_{j}W_{mjjl}n_jn_l+\frac{1}{2}\sum_{j}W_{p,mjjl}\left[n_j+n_l\right]\nonumber\\
&=\sum_{ijk}V_{mkij}\Gamma^{(2)}_{\text{c},ijkl},
\end{align}
where the first term on the right-hand side is the screened exchange and the second term the Coulomb hole. As for the SEX-like approximation, in order to get fractional occupation numbers we combine this COHSEX-like approximation with the PF functional. One can add the $\alpha$ exponent only to screened-exchange term or, to be more consistent, also to the Coulomb-hole term. Here we will consider the $\alpha$ parameter only in the SEX part. 

The COHSEX-like approximation is tested in the HEG. In this case the energy functional to be minimized reads
\bea
\frac{\mathcal{F}}{\Omega}&=&\int \frac{d \vect{k}}{(2\pi)^3} (\vect{k}^2-2\mu)n(\vect{k})\nonumber\\
&&-
\int \frac{d\vect{k}d\vect{k}'}{(2\pi)^6}v(\vect{k}-\vect{k}')
f(n(\vect{k}),n(\vect{k}'))\nonumber\\
&&+
\frac{1}{2}\int \frac{d\vect{k'}d\vect{k}}{(2\pi)^6}   W_{p}(\vect{k}')n(\vect{k})
+\mu.
\label{eqn:funcCOH}
\eea
We notice that Eq.~\eqref{eqn:funcCOH} differ from Eq.~\eqref{eqn:funcE} only by a term which does not depend of $n(\vect{k})$ ($\int d\vect{k} n(\vect{k})$ is a constant). This implies that the addition of the Coulomb hole term does not affect the optimal momentum distribution $n(\vect{k})$.

\section{Tuning the correlation in the homogeneous electron gas}
\label{Appendix:B}
\begin{figure}[b]
\centering
   \includegraphics[width=0.47\textwidth]{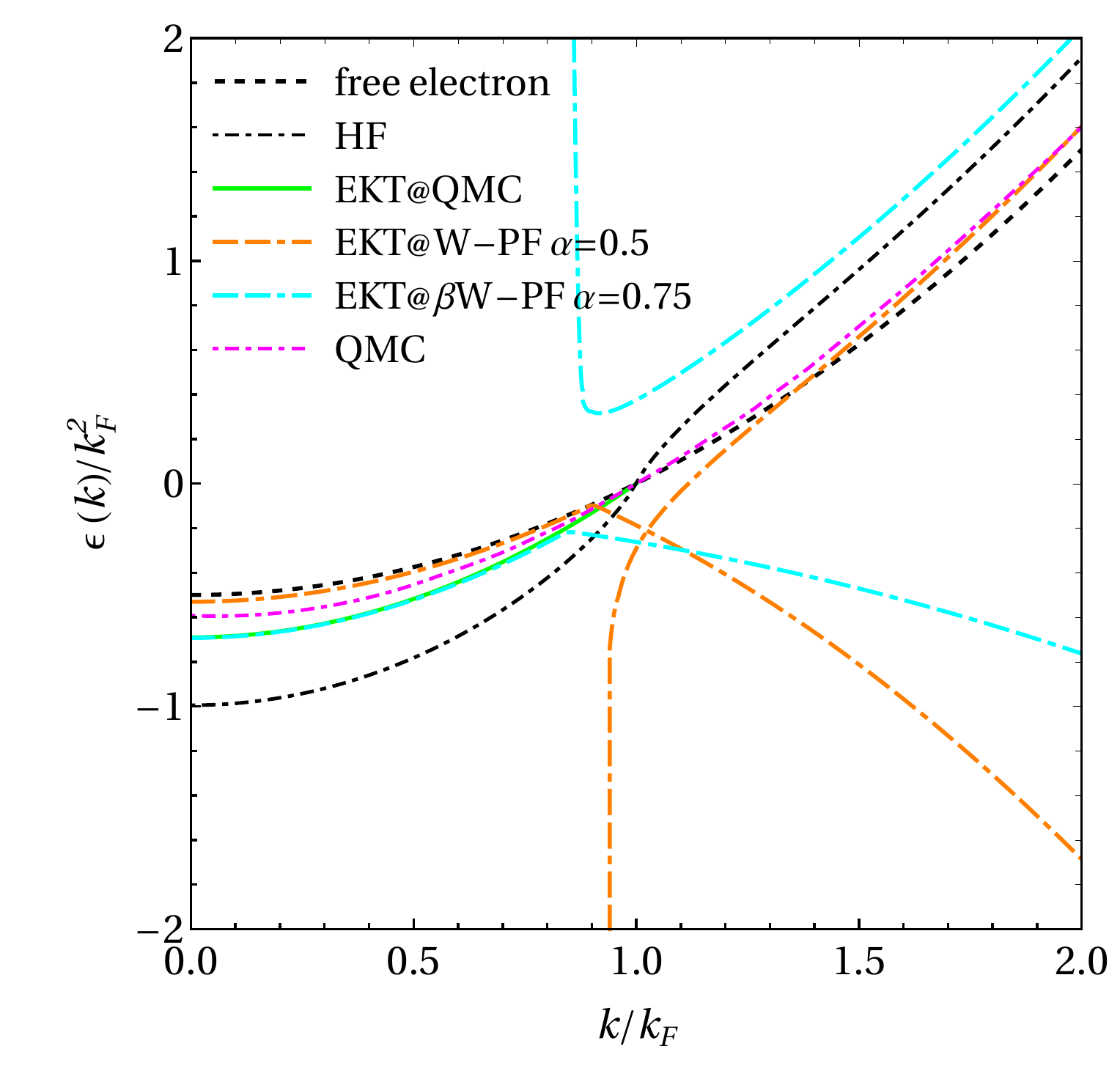}
   \vspace{-10pt}
\caption{Quasiparticle dispersion $\epsilon(k)/k_F^2$ for $r_s=3$: $\beta$\wpf is compared with \wpf.}
\label{fig:hegDispbeta}
\end{figure}

As discussed in Sec.~\ref{HEG} the \wpf suffers by an overscreening problem, which arises from double counting between $W$ and the PF. Reducing the screening and the correlation in the PF one can indeed find a quasiparticle dispersion in agreement with the EKT@QMC results. This can be shown by introducing a parameter $\beta$ in front of $W$ in the function $f$ as 

\be
f^{\beta W-\text{PF}}(n(\vect{k}),n(\vect{k}'))=\beta\frac{\bar{W}(\vect{k}-\vect{k}')}{v(\vect{k}-\vect{k}')}[n(\vect{k})n(\vect{k}')]^\alpha.
\label{Eqn:f_parametrized}
\ee
For a fixed value of $\alpha$, the parameter $\beta$ is determined in such a way to obtain the QMC correlation energy of the HEG.
The optimal value of $\alpha$ is then determined in such a way to have the same second derivative of $\epsilon^R(\vect{k})$ at $\vect{k}=0$ obtained by the EKT@QMC. We find that the optimal values of the two parameters are $\alpha=0.75$ and
$\beta=5.44$, which indicate a strong reduction of the screening. The results are reported in Fig.~\ref{fig:hegDispbeta}. One could envisage to use \eqref{Eqn:f_parametrized} with the parameters $\beta$ and $\alpha$ optimized for the HEG also for realistic systems in the same spirit as the local density approximation employed in density functional theory.

\end{document}